\documentclass[aps,showpacs,amsmath,amssymb,amsfonts,superscriptaddress]{revtex4-1}
\usepackage{times}
\usepackage{graphicx}
\usepackage{subfigure}
\usepackage{verbatim}
\usepackage{dcolumn}% Align table columns on decimal point
\usepackage{bm}% bold math
\usepackage{epsf}
\usepackage{xcolor}
\usepackage{hyperref}
\usepackage{hhline}
\usepackage{float}
\usepackage{enumerate}
\usepackage{bbm}
\usepackage{lipsum}
\usepackage{tikz,pgfplots}
\usepgfplotslibrary{fillbetween}
\usetikzlibrary{backgrounds}
\usetikzlibrary{patterns}

\definecolor{mygreen}{rgb}{0.01, 0.31, 0.59}
\definecolor{myblue}{rgb}{0.01, 0.31, 0.59}
\hypersetup{
  colorlinks=true,   % Options: false (boxed links); true (colored links).
  linkcolor=myblue,  % Color of internal links.
  citecolor=mygreen, % Color of links to bibliography.
  urlcolor =myblue    % Color of external links.
}

\definecolor{alexred}{RGB}{224,76,76}
\definecolor{alexblue}{RGB}{77,109,255}
\definecolor{alexgreen}{RGB}{108,179 45}
\definecolor{alexpurple}{RGB}{125,147,208}
\definecolor{alexred2}{RGB}{225,64,65}
\definecolor{gradred}{RGB}{238,53,69}
\definecolor{gradblue}{RGB}{130,209,248}

\newcommand{\nsth}{n_{\textrm{sth}}}
\newcommand{\msth}{m_{\textrm{sth}}}

\newcommand{\tV}{\textbf{V}}
\renewcommand{\th}{\textrm{th}}
\newcommand{\sth}{\textrm{sth}}

\renewcommand{\vr}{\rho}
\newcommand{\tr}{\textrm{tr}}

\renewcommand{\S}{\mathcal S}

\newcommand{\z}{\textbf{a}}
\newcommand{\zd}{\textbf{a}^{\dagger}}

\newcommand{\C}{\mathcal{C}}

\newcommand{\T}{\mathsf{T}}

\newcommand{\ket}[1]{| {#1} \rangle}
 
\newcommand{\ketbra}[2]{\ensuremath{|#1\rangle\!\langle#2|}}

\begin{document}

\title{Unravelling the non-classicality  role in Gaussian heat engines}

\author{A. de Oliveira Junior}
\affiliation{Faculty of Physics, Astronomy and Applied Computer Science, Jagiellonian University, 30-348 Kraków, Poland}
\author{Marcos C\'esar de Oliveira}
\affiliation{Instituto de F\'\i sica  Gleb Wataghin, Universidade Estadual de Campinas, 13083-859, Campinas, SP, Brazil}
\date{\today}

\begin{abstract} 
At the heart of quantum thermodynamics lies a fundamental question about what is genuine ``quantum'' in quantum heat engines and how to seek this quantumness, so that thermodynamical tasks could be performed more efficiently compared with classical protocols. Here, using the concept of $P$-representability, we define a function called \emph{classicality}, which quantifies the degree of non-classicality of bosonic modes. This function allows us to explore the role of non-classicality in quantum heat engines and design optimal protocols for work extraction. For two specific cycles, a quantum Otto and a generalised one, we show that non-classicality is a fundamental resource for performing thermodynamic tasks more efficiently.
\end{abstract}

\maketitle

\section{Introduction}
The effort to understand the relationship between quantum mechanics and thermodynamics makes the field of quantum thermodynamics extremely wide. Due to its broad nature, questions underlining the interplay between quantum features, such as coherence and entanglement, have been thoroughly explored using different approaches~\cite{Niedenzu_2016,Lostaglio2015, Lostaglio2015_2, Korzekwa_2016}. One aspect broadly investigated is the production of work in quantum heat engines. These are composed of one or more quantum systems that operate between two different reservoirs with the sole aim of converting heat into work. Although the laws of quantum mechanics rule these engines, this does not necessarily imply a ``quantum advantage''~\cite{10.1088/2053-2571/ab21c6,binder2019thermodynamics, Goold_2016, myers2022quantum}. In fact, a remarkable similarity was observed with classical models~\cite{Ghosh2019,GhoshM, Alicki, Gardas, Fed,Correa2014, Harbola_2012, Linden, Quan, Sourav2020}, raising the question of how the intrinsic features of quantum mechanics, such as entanglement and coherence, could be used to enhance the performance of quantum heat engines. The use of non-thermal baths, i.e., engineered reservoirs characterised by their temperatures and additional parameters, has presented a comprehensive scenario to study the relation between quantum effects and thermodynamic efficiency~\cite{Scully862,Dillenschneider_2009,PhysRevE.86.051105, Abah_2014, Niedenzu_2016, Niedenzu2018, Gonzalo1, Wang2019, Singh2020, XIAO20183051}. In particular, non-thermal baths were previously shown to be used as a resource to exceed the standard Carnot limit~\cite{Abah_2014, Abah}. In the context of heat engines, squeezed thermal baths have played an essential role, since a proof-of-principle experiment based on a nanobeam heat engine has recently been reported~\cite{PhysRevX.7.031044}. Subsequently, in~\cite{Gonzalo}, the author developed a framework underlying the main properties of this reservoir and showed that a squeezed thermal bath could be recast as a generalised equilibrium reservoir. However, in many of these works, although a non-classical feature was present, its role was not fully investigated, in favour of a previously established concept of non-passivity~\cite{Pusz1978,Lenard1978}. But how can we quantify non-classicality in heat engines? Can we quantify non-passivity through measurable quantities? An answer may be given from the principles of quantum optics.

Radiation fields with states described by specific features that can only be understood by a quantum mechanical description are known as non-classical states~\cite{dodonov2003theory, Ctlee, Marcos1}. The definition of non-classicality is not unique, but for bosonic fields, the most strict way to quantify it is through the concept of $P$-representability, which states that a given state $\rho$ is said to be $P$-representable if it can be written as a convex mixture of coherent states
\begin{equation}
\label{prepresentability}
    \rho = \int d^2 \alpha \, P(\alpha, \alpha^*) \ketbra{\alpha}{\alpha} \, , 
\end{equation}
with a proper probability distribution function $P(\alpha, \alpha^*)$ over an ensemble of states, i.e., $P(\alpha, \alpha^*)$ is non-negative and is less, or equally, singular than the delta distribution -- these states are known as \emph{classical states}. In contrast, \emph{non-classical states} corresponds to those that cannot be written as in Eq.~(\ref{prepresentability}), because its quasi-probability distribution is negative or highly singular. This definition is particularly relevant in the context of quantum information systems based on bosonic modes since only non-$P$-representable states can generate entanglement when mixed with vacuum in a beam-splitter~\cite{Kim,Xiang-bin,Nclass}. A prominent example of non-classical states is the set of squeezed states, in which the fluctuation associated with one quadrature component is below the vacuum state~\cite{scully1997quantum}. Early theoretical work in the 60s and 80s led to the conclusion that quantum fluctuations can be reduced below the shot noise in many forms of nonlinear optical interactions~\citep{nonl1,nonl2,nonl3}. For example, squeezed states are produced in nonlinear processes called degenerate parametric down-conversion, where a ``classical'' electromagnetic field drives a nonlinear medium and pairs of correlated photons of the same frequency are generated. The non-classical effects of light can be revealed in different ways. In the present case, this manifestation occurs in terms of coherence (off-diagonal elements) since the squeezing operation induces non-diagonal elements in the energy eigenbasis.

We aim to present a discussion by combining tools from quantum optics and continuous variable systems. As a result, we explore the role of non-classicality in quantum thermodynamics by quantifying the degree of $\emph{non-classicality}$ in terms of the $P$-representability. It is worth stressing that non-passivity is a necessary but not sufficient condition to detect non-classicality, and this distinction is not clear for a single-bosonic mode. However, the converse holds, i.e., all non-classical states are non-passive. For two specific examples, i.e., a quantum Otto and a generalised cycle, we show that the non-classicality is a resource for quantum heat engines, meaning that a given thermodynamic task can be performed more efficiently when this engine operates in the non-classical regime. The first analysed cycle- the Otto cycle- agrees with the results of Refs.\cite{Abah_2014, Gonzalo} where they employed a modulation in the frequency of the bosonic mode as a work parameter. Since we are interested in the trade-off between non-classicality and thermodynamic efficiency, the squeezing parameter is used as the work variable.

The paper is organised as follows. In Section~\ref{sec:Classicality_function}, we start by setting the notation and introduce some preliminaries concepts involving the Gaussian parametrisation and the definition of $P$-representability. Then, we define the classicality function and briefly recall the concept of passive states. In Section \ref{sec:Thermodynamic_setting}, the thermodynamic approach is described, and in Section \ref{sec:Quantum_heat_engines} we use the defined concepts to analyse two different cycles. Finally, we conclude with an outlook in Section~\ref{sec:Conclusion}. 

\section{Classicality and Passivity}
\label{sec:Classicality_function} 
\subsection{Classicality}
A multi-mode Gaussian state, with bosonic operators $\textbf{a} = (a_1, a^{\dagger}_1, ..., a_N, a^{\dagger}_N)^{\T}$ satisfying the canonical commutation relation~\cite{AlessioB, Cerf},
\begin{equation}
\label{CR}
    [\textbf{a}, \textbf{a}^{\dagger}] = \bigoplus_{j=1}^{n} \sigma_z, \quad \textrm{where} \quad \sigma_z = \begin{pmatrix}
    1 &0 \\ 
    0 &-1 
    \end{pmatrix} \, ,     
\end{equation}
is entirely described by its two first statistical moments, the first--order, or displacement, and the second-order moments collected in the covariance matrix (CM). For the problem under consideration, we assume a single bosonic mode with zero first moment and CM given by 
\begin{equation}
\label{CM}
    \tV = \begin{pmatrix}
    \langle a^{\dagger} a \rangle+\frac{1}{2} & -\langle a^2 \rangle \\ \\ 
    -\langle a^2 \rangle^{*}& \langle a^{\dagger} a \rangle+\frac{1}{2}
    \end{pmatrix} = \begin{pmatrix}
    \bar{n}+\frac{1}{2} & m \\ \\
    m^{*}& \bar{n}+\frac{1}{2}
    \end{pmatrix} \, .
\end{equation}
The diagonal elements of $\tV$ represent the mean photon number, or occupation, whereas the off-diagonal elements represent correlations between different eigenstates, and these will be referred to as coherence parameter. A system described by a Hamiltonian $H$, and in thermal equilibrium with a thermal environment at temperature $T$, is described by a thermal state
\begin{equation}
\label{eq:thermal}
  \rho_{\text{th}} = \frac{e^{-H/k_B T}}{Z_{\text{th}}} \quad \text{with} \quad Z_{\text{th}} = \tr[e^{-H/k_B T}] ,  
\end{equation}
where $k_B$ denotes the Boltzmann constant. This state is an example of a Gaussian state, with null first moment and covariance matrix given by  
\begin{equation}\label{thermalV}
\tV_{\th} =  \begin{pmatrix}
    \bar{n}_{\th}+\frac{1}{2} & 0 \\ \\
    0& \bar{n}_{\th}+\frac{1}{2} 
    \end{pmatrix} \, ,
\end{equation}
where the thermal average number of photons $\bar{n}_{\th} = (e^{\hbar \omega/k_B T}-1)^{-1}$ is given by the Bose-Einstein distribution. A squeezed thermal state $\vr_{\textrm{sth}}$ is a Gaussian state with zero displacement and CM equal to
\begin{equation}
\label{CMb}
    \tV_{\textrm{sth}} = \begin{pmatrix}
    \bar{n}_{\textrm{sth}}+\frac{1}{2} & m_{\textrm{sth}} \\ \\
    m^*_{\textrm{sth}}& \bar{n}_{\textrm{sth}}+\frac{1}{2}
    \end{pmatrix} =\begin{pmatrix}
    \left(\bar{n}_\th+\frac{1}{2}\right)\cosh 2r & \left(\bar{n}_\th+\frac{1}{2}\right)\sinh2r \\ \\
    \left(\bar{n}_\th+\frac{1}{2}\right)\sinh 2r& \left(\bar{n}_\th+\frac{1}{2}\right)\cosh 2r
    \end{pmatrix} \, .
\end{equation}
where now the occupation and coherence parameters are given by $\bar{n}_{\textrm{sth}} = \bar{n}_{\textrm{th}} \cosh 2r + \sinh^2 r$, with $r$ being the squeezing parameter, and $m_{\textrm{sth}} = (\bar{n}_{\textrm{th}}+1/2) \sinh 2r$, respectively. As mentioned before, without loss of generality, the direction in the phase space in which the squeezing occurs is set to zero. Therefore, a squeezed thermal state is completely parameterised by its temperature and the squeezing parameter

\begin{figure}
\centering
\includegraphics{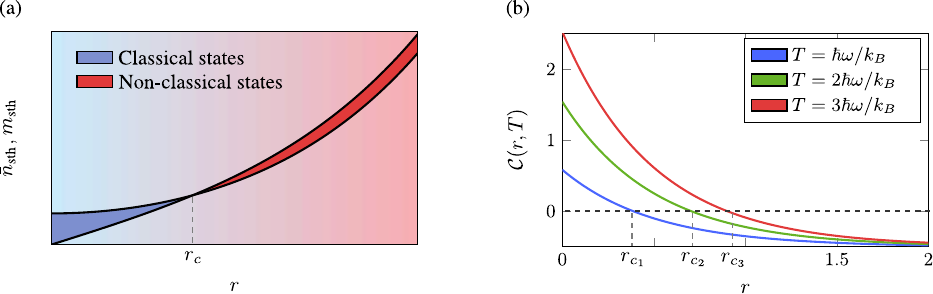}
    \caption{(a) A sketch of the parameters $\bar{n}_{\textrm{sth}}$ and $m_{\textrm{sth}}$ as a function of the squeezing parameter. Initially, the system is prepared in a thermal state at temperature $T$ and is classical (blue region). By applying a squeezing operation and modulating the squeezing parameter, the system goes from a classical to a non-classical state (red region). This transition occurs with respect to a critical squeezing parameter $r_c(T)$. In (b) we plotted the classicality as a function of the squeezing parameter for three different temperatures. Observe that after $r_c$ the system becomes non-classical, thereby characterising a negative classicality function $\C <0$}
    \label{Fig_classicalityofasystem}
\end{figure}

By definition, a state is $P$-representable (or classical) if its covariance matrix satisfies the following relation~\cite{Englert} (see Appendix~\ref{appendix} for more details).
\begin{equation}
\label{prep}
    \tV -\frac{\mathbbm{1}}{2} \geq 0 \, ,    
\end{equation}
meaning explicitly from Eq.~\eqref{CM} that 
\begin{equation}
\label{Condition}
    \bar{n} > |m| \, .   
\end{equation}
From the condition (\ref{Condition}), it is inferred that not all states are classical and their classicality depends on the occupation and coherence parameter. Although the thermal state is always classical ($m$ is always zero), a squeezed thermal state may not be depending on $r$. Increasing the squeezing parameter, a transition from the classical to the non-classical regime is observed at a critical value $r \equiv r_c$. This threshold, $r_c$, is reached when $\bar{n}_{\textrm{sth}} = m_{\textrm{sth}}$. Consequently, for a fixed temperature, the critical squeezing parameter is given by
\begin{equation}
r_c(T) = \frac{1}{2}\ln \left[\coth \left(\frac{\hbar \omega}{2k_B T} \right) \right] . 
\end{equation}
Physically, $r_c$ corresponds exactly when one of the quadratures reaches the uncertainty bound, and the state cannot be represented as being merely a mixture of coherent states. In Fig.~\ref{Fig_classicalityofasystem}(a), for a fixed $T$, we plot the behaviour of $\nsth$ and $\msth$ as a function of the squeezing parameter. The states between these two curves are classical (blue-shaded area). For any $r > r_c$, the coherence parameter is larger than the mean photon number $\msth > \nsth$, and therefore the state is non-classical (red-shaded area). 

Following the $P$-representability criteria~\eqref{prep}-\eqref{Condition}, we can define the \emph{classicality} function as the difference between the mean number of photons and the coherence parameter: $\C\equiv\bar{n}-|m|$. This function not only makes the Eq~.\eqref{Condition} operational but also gives a non-classicality quantifier. Hence, $\C<0$ indicates that the system is described by a non-classical state. For the squeezed thermal states, the classicality function is given by
\begin{align}
\label{Classicalityfunction}
\begin{split}
    \C(r, T):= \frac{1}{2}\coth\left(\frac{\hbar \omega}{2 k_B T} \right)e^{-2r} - \frac{1}{2}  \, .
\end{split}
\end{align}
Observe that the classicality of a given state depends exclusively on the temperature and the squeezing parameter. Classical states with different temperatures will typically be characterised by different classicalities. As the temperature increases, the system becomes more classical, and a more considerable value of $r$ is required to achieve its non-classical character (see Fig.~\ref{Fig_classicalityofasystem}(b)). As previously mentioned, this is related to the uncertainty in the quadratures of the bosonic mode -- a non-classical state is attained when this uncertainty falls below the shot noise. 

Note that coherent states also present non-diagonal terms in the covariance matrix, which could also be identified as a source of non-classicality. However, these states do not show any non-classical manifestation, and this is clear from the definition of the classicality function since, for coherence states, its classicality function is zero. Therefore, coherent states can be considered to be on the verge of being classical and non-classical states.

\subsection{Passivity}

Consider a quantum system described by a Hamiltonian $H$ and prepared in a state $\rho$. We say that $\rho$ is passive if
\begin{equation}
\tr(H U \rho U^{\dagger}) \geq \tr(H\rho) \, ,
\end{equation}
for all unitaries $U$ that act on the system. States satisfying the above equation cannot have their energy reduced by deterministic unitary transformations. The necessary and sufficient conditions for a state to be passive require that the former commute with the Hamiltonian, i.e., $[\rho, H] = 0$, and for the eigenvalue decomposition $\rho=\sum_{n=0}^{d-1}p_n |n\rangle\langle n|$, we find that $\epsilon_n \leq \epsilon_m$ implies $p_n \geq p_m$ for all $n$ and $m$ in $\{1, ..., d\}$~\cite{koukoulekidis2021geometry}. Moreover, a state is $k$-passive if $\rho^{\otimes k}$, with $k \in \mathbb{N}$, is passive. If for all integers $k \geq 1$, the state is $k$-passive, thus it is called completely passive.

A thermal state, given by a diagonal covariance matrix, is the only completely passive state. This means that it does not mind how many copies of a thermal state we have, unitary operations cannot lower its energy. Consequently, any resource state for a cyclic engine must be out of thermal equilibrium~\cite{brown2016passivity} and any $m \neq 0$ in Eq.~\eqref{CM} defines a non-passive state, but does not necessarily a non-classical state, as we see in Fig.~\ref{Fig_classicalityofasystem}(b). 
For example, a bosonic mode in contact with two reservoirs in thermal states at temperature $T_1=T_2$ is passive and, to produce work, it is necessary that the reservoirs are out of equilibrium, $T_1\neq T_2$. Alternatively, by squeezing the mode, the degree of non-passivity will increase, and also its non-classsicality, which is measured by Eq.~\eqref{Classicalityfunction}. However, the state is only non-passive and non-classical for $r > r_c$. This distinction will be crucial during the description of the two thermodynamic protocols.

\begin{figure*}
\includegraphics{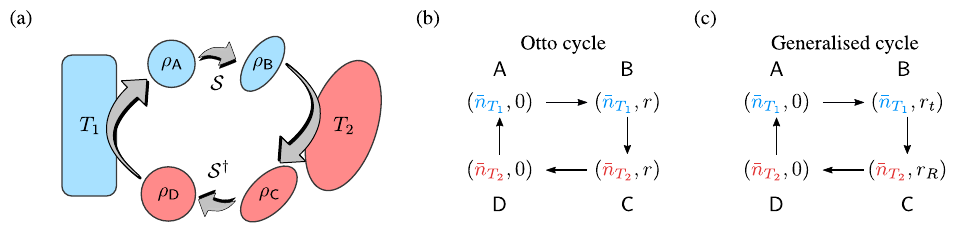}
\caption{(a) Schematic diagram of the four-step protocol in which the bosonic mode is coupled with two different reservoirs. The unitaries $\S$ and $\S^{\dagger}$ represent the adiabatic squeezing and unsqueezing modulation of the bosonic mode. Diagrammatic representation of the (b) Otto and (c) generalised cycles. While in the former, the classicality varies during the system's interaction with the hot reservoir, in the latter, it is kept constant.}
\label{Fig_protocol}
\end{figure*}

\section{Thermodynamic setting}
\label{sec:Thermodynamic_setting}
We consider a quantum heat engine based on a single bosonic mode as a working substance. The bosonic mode operates between two reservoirs at different temperatures: a cold thermal bath at temperature $T_1$ and a hot squeezed thermal bath at $T_2 > T_1$ with a squeezing parameter $r_R$. The bosonic mode undergoes two different cycles, with the work parameter being the squeezing. More precisely, both processes consist of reversible operations of squeezing and unsqueezing the mode through different protocols designed to explore the role of the classicality function (see Fig.~\ref{Fig_protocol}(a)).

\subsection{System-bath interaction}

We address the problem by assuming a weak coupling regime between the single-mode and heat baths, where the dynamics is modelled adopting a time-independent Lindblad master equation $\dot{\rho} = \mathcal{L}(\rho)$. The single-mode, initially isolated and described by a Hamiltonian $H = \hbar \omega_0 a^{\dagger} a$, is weakly coupled to a bosonic reservoir, $H_B = \sum_j \hbar \omega_j b^{\dagger}_j b_j$, via a coupling constant $k_j$ in the rotating wave approximation according to the interaction Hamiltonian $\mathcal{V} = \sum_j \hbar(k_j^* a b^{\dagger}_j+k_j a^{\dagger} b_j)$. The cold reservoir is prepared in a thermal state, and the hot reservoir is prepared in a squeezed thermal state, i.e., a thermal state at temperature $T_2$ to which the squeezing operator has been applied to each oscillator composing the bath. Under the Born-Markov approximation, the master equation for the single mode when interacting with the squeezed thermal bath is given by~\cite{scully1997quantum, breuer2002theory}:
\begin{eqnarray}
\label{masterequationgenerall}
    \dot{\rho} &=& \gamma(\bar{n}_{\textrm{sth}}+1)\left[a\rho a^{\dagger}-\frac{1}{2}\{a^{\dagger} a, \rho\}\right]+\gamma \bar{n}_{\textrm{sth}}\left[a^{\dagger} \rho a-\frac{1}{2}\{a a^{\dagger}, \rho\}\right] \nonumber+\gamma m_{\textrm{sth}}\left[a^{\dagger} \rho a^{\dagger} - \frac{1}{2}\{a^{\dagger}{^2},\rho\}\right]+\gamma m_{\textrm{sth}}^*\left[a \rho a- \frac{1}{2}\{a^2,\rho\}\right],
\end{eqnarray}
where $\gamma$ is the coupling constant between the single mode and the bath. The parameters $\bar{n}_{\textrm{sth}} =\bar{n}_{\textrm{th}} \cosh 2r + \sinh^2 r$ and $m_{\textrm{sth}} = (\bar{n}_{\textrm{th}}+1/2) \sinh 2r$ are the same as those in Eq.~\eqref{CMb}. The stationary solution of Eq.~(\ref{masterequationgenerall}) results in the squeezed thermal state~\cite{Gonzalo}, 
\begin{equation}
\label{eq:stationarystate}
    \rho_{\textrm{sth}} = \frac{e^{-\beta_{\textrm{s}}(H- \mu \mathcal{A})}}{Z_{\textrm{sth}}} \quad \text{with} \quad Z_{\textrm{sth}} =\tr[e^{-\beta_{\textrm{sth}}(H- \mu \mathcal{A})}] ,
\end{equation}
where $\beta_{\textrm{sth}} = \beta \cosh (2r)$ the generalised inverse temperature, and $\mu = \tanh(2r)$ the chemical-like potential. The operator $\mathcal{A}$ has the following form
\begin{equation}
    \mathcal{A} = -\frac{\hbar \omega}{2}(a^{\dagger}{^2}+a{^2}),  
\end{equation}
and it is known as the second-order moment's asymmetry since it measures how asymmetric, or compressed, is the state in the phase-space picture. In this work, we will be interested in the regime in which the interaction time between the system and the reservoir goes to infinity, i.e., the asymptotic limit. However, the finite and non-equilibrium regime can also be explored by considering the complete dynamical evolution in Eq.~(\ref{masterequationgenerall}).

\subsection{Thermodynamic work}

The next step in our description is to define work. Moving from the macroscopic description to the microscopic quantum realm, a clear picture of what is work and heat becomes blurred, as fluctuations and randomness are fundamentally unavoidable. Thus, carrying out these concepts into the quantum realm is not a straightforward task. However, developments in stochastic thermodynamics~\cite{talkner2016aspects,JARZYNSKI2007495}, open quantum systems~\cite{Alicki,cuzminschi2021extractable}, and quantum information theory~\cite{horodecki2013fundamental,brandao2013resource} have led to significant progress in these efforts.

In a complete analogy with classical thermodynamics, we define work as a controllable (coherent) energy exchange related to a parameter of the system and heat as an (incoherent) flux of energy that cannot be given or subtracted in a controlled manner (or be helpful in some process). As we are interested in a single bosonic mode prepared in a squeezed thermal state with mean energy equal to 
\begin{equation}
\label{interalenergy}
    E(r,\bar{n}_{\textrm{th}}) = \hbar \omega\left(\bar{n}_{\textrm{th}}+\frac{1}{2} \right) \cosh2r \, ,
\end{equation} 
one can differentiate Eq.~\eqref{interalenergy} to obtain two different contributions to the mean energy
\begin{align}
\label{differentiation}
    dE &= \left(\frac{\partial E}{\partial r}\right)_{\bar{n}_{\th}}\textup{d}r+\left(\frac{\partial E}{\partial \bar{n}_{\th}}\right)_{r}\textup{d}\bar{n}_{\th} = \left[ 2\hbar \omega \left(\bar{n}_{\th}+\frac{1}{2} \right) \sinh 2r \right]\textup{d}r+ (\hbar \omega \cosh 2r) \textup{d} \bar{n}_{\th} \, ,
\end{align} 
where we identify the first term as work (an external control parameter) and the second term as heat
\begin{align}
\label{workandheat}
    \delta W = -2\hbar \omega \left(\bar{n}_{\textrm{th}}+\frac{1}{2} \right)\sinh 2r \, \textup{d}r \quad \text{and} \quad \delta Q = \hbar \omega \cosh 2r \, \textup{d} \bar{n}_{\textrm{th}} \, .
\end{align}
The minus signal in the work expression is adopted because, during the preparation of the squeezed state, external energy is required.

The connection between the extracted work and non-classicallity (coherence) is made using the relative entropy of coherence~\cite{Plenio}, which reveals that the amount of coherence present on the Fock basis is proportional to $m$ and inversely proportional to $T$
\begin{equation}
    C(\rho_{\sth}) \approx  \beta_{\textrm{s}} m \, .  
\end{equation}
where $C(\rho) = S(\rho_{\textrm{diag}})- S(\rho)$ is the relative entropy of coherence, $S(\rho):= \text{Tr}(\rho \log \rho)$ is the von-Neumann entropy, and $\rho_{\textrm{diag}}$ denotes the state obtained from $\rho$ by removing all off-diagonal elements. For an isothermal process, the non-equilibrium free energy characterises the optimal amount of work that can be extracted from a given system with the help of a thermal reservoir~\cite{Esposito, Parrondo2015}. This result also holds for a squeezed thermal bath~\cite{Gonzalo}. In particular, we can show that the work extracted in this case is directly related to $m$ and is given by
\begin{equation}
     W = \hbar \omega \tanh(2r) \Delta m \, .
\end{equation}
Therefore, by squeezing a bosonic mode, coherence between energy eigenstate is created, and this can be quantified by the parameter $m$. In our description, the presence of non-diagonal elements in the covariance matrix and a negative classicality function will be an indicator of non-classicality. Furthermore, the amount of non-classicality present in a given bosonic system can be quantified by the classicality function~\eqref{Classicalityfunction}.

\section{Quantum heat engines}
\label{sec:Quantum_heat_engines}
We now turn our attention to investigating the performance of a heat engine that operates in two different cycles. The first is an Otto cycle with isentropic strokes corresponding to a modulation of the squeezing parameter. In this cycle, the classicality of the system increases during its interaction with the squeezed thermal bath since the work parameter (squeezing) is kept constant and the temperature increases. To understand how the degree of non-classicality affects the performance of the heat engine, we will also consider a second cycle, where instead of keeping a fixed squeezing parameter during the isothermal process, the classicality is maintained constant so that we can analyse the trade-off between the degree of non-classicality and the efficiency of the thermodynamic protocol. 

\subsection{\label{sec:lin} Otto Cycle}
The Otto cycle consists of four consecutive strokes, two isochoric and two isentropic processes, as shown in Fig.~\ref{Fig_protocol}(b). During the isentropic stroke, an external agent modulates the squeezing parameter of the harmonic oscillator between $0$ and $r$. Heating and cooling results from the coupling with the two heat baths at temperatures $T_1 = \hbar \omega/k_B$ and $T_2=2\hbar \omega/k_B$. The hot squeezed thermal bath is prepared with the same squeezing parameter as the system after the first stroke. The state of the working fluid that performs the cycle is denoted by $\rho_{\mathsf{A}}$, $\rho_{\mathsf{B}}$, $\rho_{\mathsf{C}}$, and $\rho_{\mathsf{D}}$, respectively. Furthermore, the mean number of photons will be denoted by $\bar{n}_{T_i} = (e^{\hbar \omega/k_B T_i}-1)^{-1}$, where $i = 1,2$.
\begin{figure*}
\includegraphics{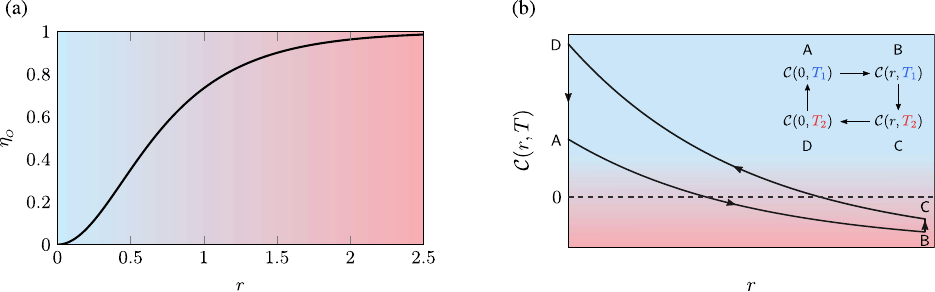}
\caption{(a) Efficiency, $\eta_o$, of the Otto heat engine as a function of the squeezing parameter. The efficiency does not depend on the temperature, but only on the squeezing parameter. The difference in colour indicates an increasing degree of non-classicality with $r$. In (b) we have the classicality function (\ref{Classicalityfunction}) in the Otto cycle as a function of the squeezing parameter for $T_1 = \hbar\omega/k_B$ and $T_2 = 2\hbar\omega/k_B$. In the inset, we present the diagrammatic representation of the Classicality function over the cycle. The blue and red-shaded denotes the classical, and non-classical regime, respectively.}
\label{Fig_OttoandEf}
\end{figure*}

The cycle starts with the working fluid in equilibrium with the cold thermal reservoir. During the first stroke, the working fluid is decoupled from the reservoir and a unitary squeezing operation $\S $ is applied to the system, resulting in isentropic compression. The work done in the single mode reads
\begin{align}
\begin{split}
\label{workAB}
    W_{\mathsf{AB}} = -2\hbar \omega \left ( \bar{n}_{T_1} +\frac{1}{2}  \right )\sinh^2 r \, \, ,
\end{split}
\end{align} 
where $r$ is the squeezing parameter of the transformation. In the second stroke, the bosonic mode is put in contact with the squeezed thermal bath, and the squeezing parameter remains constant, resulting in an ``isochoric'' process. Then, the mode relaxes to the steady state $\rho_{\mathsf{C}}$. In this step, the heat exchanged between the system and the squeezed thermal bath is
\begin{align}
\label{aheat}
    Q_{\mathsf{BC}} = \hbar \omega (\bar{n}_{T_2} -\bar{n}_{T_1}) \cosh 2r  \, .    
\end{align}
During the third stroke, the working fluid is again detached from the reservoir, and a unitary unsqueezing operation is applied to the mode $\S^{\dagger}$, bringing its squeezing parameter back to $0$. Consequently, the amount of work extracted in this stroke is given by
\begin{equation}
\label{workBC}
    W_{\mathsf{CD}} = 2\hbar \omega \left ( \bar{n}_{T_2}+\frac{1}{2}  \right )\sinh^2 r \, .
\end{equation}
Finally, the cycle is closed by bringing the bosonic mode into contact with the cold thermal bath at temperature $T_1$, and consequently the state relaxes to $\rho_{\mathsf{A}}$. During the last isochoric process, the heat transferred from the cold reservoir to the system yields the following results
\begin{equation}
\label{rheat}
    Q_{\mathsf{DA}} = \hbar \omega (\bar{n}_{T_1}-\bar{n}_{T_2}) \, .    
\end{equation}

The work over the cycle is given by the two isentropic contributions
\begin{equation}
\label{totalwork}
    W_{\textrm{cycle}} = 2 \hbar \omega \left(\bar{n}_{T_2}-\bar{n}_{T_1}\right)\sinh^2 r \,,
\end{equation} 
and the efficiency of the engine, defined as the ratio between the output work and the heat absorbed from the hot reservoir, can be obtained from the previous results. Specifically Eqs.(\ref{aheat}) and (\ref{totalwork}) allow us to compute the efficiency  
\begin{equation}
\label{efficiency}
    \eta_{o} = 1 -\frac{1}{\cosh 2r} \, .
\end{equation}

As already expected from Ref.~\cite{Abah_2014}, the efficiency increases with the squeezing parameter (see Fig.~\ref{Fig_OttoandEf}~(a)) and for higher values of squeezing, the efficiency will approach unity, but will never exceed it. Note that, compared to the classical case, the efficiency does not explicitly depend on the temperature of the reservoirs.

The behaviour of the classicality function throughout the cycle is plotted in Fig.~\ref{Fig_OttoandEf}~(b). Observe that in the first stroke ($\mathsf{AB}$), non-classicality is added to the single mode by the action of the squeezing operator. In other words, a classical state with no coherence was taken to the non-classical regime (red region), now characterised by a non-classical state with coherence in its energy eigenbasis.

In the next stroke ($\mathsf{BC}$), the squeezing parameter stays constant, while the temperature of the bosonic mode increases, meaning an increase in the classicality function. This can be interpreted as if the bath had consumed part of the non-classicality that was added in the first stroke. The last two strokes ($\mathsf{CD}$) and ($\mathsf{DA}$) correspond to the unsqueezing and the lowering of the temperature of the bosonic mode. Thus, in this step, the classicality function increases until its maximum and reaches its initial value. From Fig.~\ref{Fig_OttoandEf}~(b), we see that non-classicality is an essential resource in thermodynamics as the efficiency of the protocol increases with the degree of squeezing. More precisely, with the degree of non-classicality. 

Note that this protocol does not fully explore the non-classicality added in the bosonic mode as the interaction with the hot reservoir causes an increase in the system's classicality. This motivates us to introduce a novel stroke type in which the degree of classicality is kept constant during the system-reservoir interaction. 

\subsection{Generalized cycle}
\label{sec:GC}
We consider the same setup as before: a quantum heat engine operating between a cold thermal bath at temperature $T_1 = \hbar \omega/k_B$, and a hot squeezed thermal reservoir at $T_2 = 2 \hbar \omega/k_B$, with squeezing parameter $r_R$ (see Fig.~\ref{Fig_protocol} (c)). The main difference is that now we want to preserve the same degree of non-classicality added during the interaction of the system with the hot reservoir: this implies that the squeezed thermal bath must be prepared with a different squeezing parameter than the system; essentially, because the system will absorb heat from the hot bath, so its temperature and classicality increases. Consequently, to compensate for the addition of classicality, the system must have its squeezing parameter modulated from $r_t$ to $r_R$, where $r_t$ is the squeezing parameter related to the first stroke. Consequently, during the interaction between the system and the reservoir, we impose the following condition: 
\begin{equation}
\label{conditionclassicality}
    \mathcal{C}(r_t, \bar{n}_{T_1}) = \mathcal{C}(r_R, \bar{n}_{T_2}) \, ,
\end{equation}
The cycle starts with the working fluid in thermal equilibrium with the cold thermal bath, and the procedure followed in the first stroke is the same as in the Otto cycle: the working fluid is disconnected from the cold bath and its squeezing parameter modulated from 0 to $r_t$. In the second stroke, the bosonic mode is put in contact with the squeezed thermal bath. To keep the classicality constant, the stationary state of the system is slowly varied, so that the condition~\eqref{conditionclassicality} is satisfied. According to the definitions~\eqref{workandheat}, in the second stroke, work, and heat can be obtained from
\begin{align}
\label{workandheat2}
    W_{\mathsf{BC}} = 2\hbar \omega \int_{r_t}^{r_R}\left(\bar{n}_T+\frac{1}{2} \right)\sinh 2r \, \textup{d}r  \quad , \quad Q_{\mathsf{BC}} = \hbar \omega \int_{r_t}^{r_R} \cosh 2r \, \textup{d} \bar{n}_T 
\end{align} 
Moreover, to satisfy condition~\eqref{conditionclassicality}, work is performed on the system during its interaction with the squeezed thermal bath. The relation between $r_t$ and $r_R$, can be obtained using Eq.~\eqref{conditionclassicality} and gives
\begin{align}
\begin{split}
    r_R = r_t+\frac{1}{2}\ln \left(\frac{\bar{n}_{T_2}+\frac{1}{2}}{\bar{n}_{T_1}+\frac{1}{2}} \right).  
\end{split}    
\end{align}
In the third stroke $\mathsf{BC}$, the system is again detached from the reservoirs and an unsqueezing operation, $\mathcal{S}^{\dagger}$, is applied to the mode, changing its squeezing parameter adiabatically back to 0. The cycle is closed by putting the bosonic mode in contact with the cold thermal reservoir and relaxing back to $\rho_{\mathsf{A}}$.

\begin{figure*}
\includegraphics{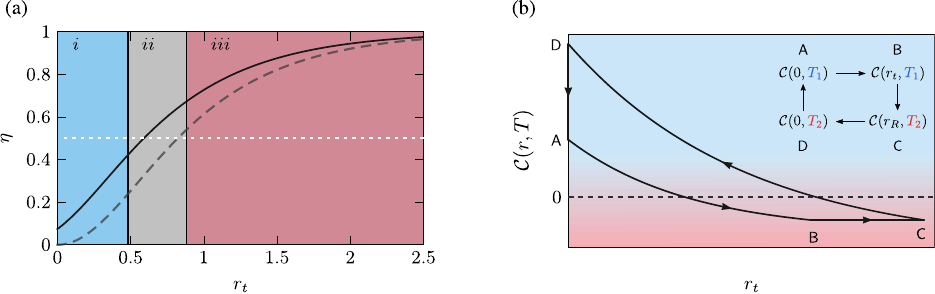}
\caption{(a) The thermodynamic efficiency of the generalised cycle (black curve), Otto cycle (dashed grey) curve and the Carnot efficiency (dashed white) as a function of the squeezing parameter for $T_1 = \hbar \omega/k_B$ and $T_2 = 2\hbar \omega/k_B$. Also illustrated, a phase diagram with the three regimes of the cycle \emph{(i, ii, and iii)}. The three different colours denote the degree of non-classicality of the bosonic mode. While blue and red indicate the classical, and non-classical behaviour of the mode, the grey colour represents that the mode is non-classical for $T_1$, but classical for $T_2$. In (b) the Classicality function (\ref{Classicalityfunction}) in the generalised cycle as a function of the squeezing parameter for $T_1$ and $T_2$. The blue and red-shaded colours denotes the classical, and non-classical regime, respectively.}
\label{Fig_GeneralisedandEf}
\end{figure*}

Once the work and heat are calculated for each stroke, the efficiency of the cycle is determined. Differently from the first protocol, now the efficiency depends on the temperatures and does not have a straightforward expression as before,
\begin{equation}
    \eta_G = 1-\frac{4 e^{2 r_t} \left[\coth \left(\frac{\hbar \omega }{2 T_2 k_B}\right)-\coth \left(\frac{\hbar \omega }{2 T_1 k_B}\right)\right]}{e^{4 r_t} \tanh \left(\frac{\hbar \omega }{2 T_1 k_B}\right) \coth ^2\left(\frac{\hbar \omega }{2 T_2 k_B}\right)-\coth \left(\frac{\hbar \omega }{2 T_1 k_B}\right) \left\{e^{4 r_t}-2 \log \left[\tanh \left(\frac{\hbar \omega }{2 T_1 k_B}\right) \coth \left(\frac{\hbar \omega }{2 T_2 k_B}\right)\right]\right\}} \, .
\end{equation}

A comparison between the efficiency of both cycles is shown in Fig.~\eqref{Fig_GeneralisedandEf}(a), where the generalised cycle is represented by a black line, and the Otto cycle by a dashed grey line. The generalised cycle is always more efficient than the Otto cycle, since the classicality of the system is kept constant along the third stroke. Consequently, work must be done on the system to satisfy the condition~\eqref{conditionclassicality}, which means higher extraction of work in the third stroke. Note that in the Otto cycle, the hot bath consumes part of the non-classicality that is given to the bosonic mode. Here, we prevent this by imposing the condition~\eqref{conditionclassicality}. 

As can be observed in Fig.~\eqref{Fig_GeneralisedandEf}(a), the non-classical character of the system depends on its temperature and squeezing parameter. Consequently, for a single-mode operating at two different temperatures, there are two different critical squeezing parameters, denoted by $r_{c_1}$ and $r_{c_2}$. This implies that this cycle can be investigated in three different regimes. That is, when the state of the single-mode is (as summarised in the phase diagram of Fig.~\ref{Fig_GeneralisedandEf}(a)):
\begin{enumerate}[i.]
    \item \emph{Classical for both temperatures.}
    \item \emph{Non-classical for $T_1$, but classical for $T_2$.}
    \item \emph{Non-classical for both temperatures.}
\end{enumerate}

\emph{Region (i)} corresponds to the case where the system is classical, $\mathcal{C}(r,T) \geq 0$, but non-passive and only the degree of non-passivity plays a role. Although there is squeezing, the uncertainty in both quadratures is above the shot-noise limit, and the squeezing can be seen as a modulation of the single-mode frequency from $\omega_1$ to $\omega_2$, where $\omega_2 > \omega_1$, in a similar fashion to previous works.

\emph{Region (ii)} corresponds to the case where the state of the system is non-passive and non-classical for $T_1$, but classical for $T_2$. Compared to \emph{Region (i)}, the non-passivity combined with the non-classicality of the state increases the efficiency of the protocol. Since the non-classical behaviour depends on the external temperature, the single mode undergoes two transitions during the cycle: classical to non-classical (first stroke) and non-classical to classical (third stroke) due to the temperature increasing (see~Fig.\ref{Fig_GeneralisedandEf} (b)). Observe that Carnot's efficiency (dashed grey line) is surpassed in \emph{region (ii)}, where the working fluid starts the cycle with a non-classical character. Our understanding is that the non-classicality added is used to perform the cycle more efficiently and helps to beat Carnot's limit. However, this comparison is not totally appropriate since the problem in consideration is formulated in a different context from which Carnot's cycle was proposed.

Finally, \emph{region (iii)} is the most efficient. In this particular case, the state of the system is non-passive and non-classical for both temperatures. As a matter of comparison, the classical Carnot efficiency is also plotted (see~Fig.\ref{Fig_GeneralisedandEf}). As we can see in Fig.~\eqref{Fig_GeneralisedandEf}(a), with increasing squeezing parameter, the efficiencies in the Otto (grey dashed line) and generalised cycle (black line) become closer to each other. This happens because the absorbed heat increases with the squeezing parameter, while the rejected heat does not. Therefore, for higher values of squeezing ($r >> r_{c_2}$), the difference between the two cycles becomes tight. In other words, the degree of non-classicality is so high that the type of cycle does not affect the performance of the quantum heat engine.

It is worth mentioning that in both cycles, we see a monotonic increase in efficiency with the squeezing parameter. There is no visible change in the respective curve when the squeezing parameter reaches the critical value corresponding to the transition from a classical to a non-classical regime. The lack of signature occurs because we are dealing with two different temperatures, and our definition of classicality is formulated for a single temperature. Moreover, the efficiency of the heat engine takes into account all four strokes, and in each, we are interested in the energy exchange. Consequently, the classical or non-classical character of the working fluid contributes only to the average energy, so a discontinuity or a signature discriminating between these two regimes, or their change, is not expected. 

\section{Conclusion}
\label{sec:Conclusion}
A central aim in quantum thermodynamics is the search for quantum advantages in a given thermodynamical task. Here, we have seen an example of a quantum advantage using a simple measure of classicality for squeezed states. This work presents an analysis of how non-classical features of light will play a role in thermodynamics from a $P$-representability perspective. Specifically, squeezed states have been shown to be a resource for thermodynamics. This, in turn, allows us to make use of the framework developed in~\cite{Abah_2014, Gonzalo} to track the trade-off between thermodynamic efficiency and the non-classicality presented in the work substance. The thermodynamic implications of squeezed states on heat engine performance have been studied in numerous previous works, especially in the case of the harmonic quantum Otto engine. However, the external control parameter modified during the isentropic stroke is the frequency, whereas here we is the squeezing parameter.  In some sense they are equivalent, as the variation of squeezing can be understood as a variation of frequency, as the energy term $E= \hbar\omega (\bar{n}_\textrm{sth}+1/2) =  \hbar\omega(\bar{n}_{\textrm{th}}+1/2) \cosh 2r$, can be recast as the energy of a harmonic oscillator, $E= \hbar\omega'(r) (\bar{n}_\textrm{sth}+1/2)$, with $ \omega'(r)=\omega \cosh 2r$, a squeezing dependent frequency. For obvious reasons the same is not true about the coherences $m_\textrm{sth}$, as they are absent for the harmonic oscillator. Our treatment allowed us to identify a novel stroke type in which the degree of classicality is held constant, allowing for the introduction of a new class of generalised heat engine cycles.

In this study, we provided a simple expression, namely the classicality function, which allows us to distinguish classical and non-classical states. Since the squeezing parameter is the route to non-classicality, we treated it as a working parameter. Then, two different cycles where the classicality function is explored were analysed. Our main result shows that the efficiency of a quantum heat engine is enhanced when it operates in the non-classical regime, whereas it is less efficient in the classical one. An important point to be mentioned is the role of passivity/non-passivity. In our description, it can be straightforwardly observed that the anti-squeezing operation $\S$ induces maximal work extraction, i.e., ergotropy. However, here we have the interplay between non-passivity and non-classicality. The discussion of the generalised cycle was a way to circumvent this issue and to explicitly show that the degree of non-classicality is indeed a resource for quantum heat engines. 

Our work opens many potential avenues for future research. First, in the current work, we have focused exclusively on a single mode as work substance, so a natural question concerns its extension for a multi-mode scenario where entanglement can also be explored. Moreover, since our discussion is only in terms of the equilibrium regime, this manuscript leaves open the possibility of a significant amount of follow-up research. Finite-time analysis (either in the endoreversible or fully nonequilibrium regime) would open up the possibility of studying the consequences of the degree of classicality on a wide range of other engine characterisations, including power output and efficiency at maximum power. As shown in~\cite{PhysRevX.7.031044}, quantum engines employing squeezed states are also experimentally accessible, making experimental implementations of this work a near-term possibility.

\begin{acknowledgments}
The authors thank Juan Parrondo, Gonzalo Manzano, Gabriel Landi, Marcus Bonan\c ca and Kamil Korzekwa for helpful discussions. AOJ acknowledges financial support by the Foundation for Polish Science through the TEAM-NET project (contract no. POIR.04.04.00-00-17C1/18-00). MCO acknowledges the support from the Institute of Physics \emph{Gleb Wataghin}, the financial support from CNPq (Brazil) and from the Brazilian funding agencies Conselho Nacional de Desenvolvimento Científico e Tecnológico (CNPq) and Coordenação de Aperfeiçoamento de Pessoal de Nível Superior (CAPES).
\end{acknowledgments}

\bibliography{referencias}

\begin{thebibliography}{59}%
\makeatletter
\providecommand \@ifxundefined [1]{%
 \@ifx{#1\undefined}
}%
\providecommand \@ifnum [1]{%
 \ifnum #1\expandafter \@firstoftwo
 \else \expandafter \@secondoftwo
 \fi
}%
\providecommand \@ifx [1]{%
 \ifx #1\expandafter \@firstoftwo
 \else \expandafter \@secondoftwo
 \fi
}%
\providecommand \natexlab [1]{#1}%
\providecommand \enquote  [1]{``#1''}%
\providecommand \bibnamefont  [1]{#1}%
\providecommand \bibfnamefont [1]{#1}%
\providecommand \citenamefont [1]{#1}%
\providecommand \href@noop [0]{\@secondoftwo}%
\providecommand \href [0]{\begingroup \@sanitize@url \@href}%
\providecommand \@href[1]{\@@startlink{#1}\@@href}%
\providecommand \@@href[1]{\endgroup#1\@@endlink}%
\providecommand \@sanitize@url [0]{\catcode `\\12\catcode `\$12\catcode
  `\&12\catcode `\#12\catcode `\^12\catcode `\_12\catcode `\%12\relax}%
\providecommand \@@startlink[1]{}%
\providecommand \@@endlink[0]{}%
\providecommand \url  [0]{\begingroup\@sanitize@url \@url }%
\providecommand \@url [1]{\endgroup\@href {#1}{\urlprefix }}%
\providecommand \urlprefix  [0]{URL }%
\providecommand \Eprint [0]{\href }%
\providecommand \doibase [0]{http://dx.doi.org/}%
\providecommand \selectlanguage [0]{\@gobble}%
\providecommand \bibinfo  [0]{\@secondoftwo}%
\providecommand \bibfield  [0]{\@secondoftwo}%
\providecommand \translation [1]{[#1]}%
\providecommand \BibitemOpen [0]{}%
\providecommand \bibitemStop [0]{}%
\providecommand \bibitemNoStop [0]{.\EOS\space}%
\providecommand \EOS [0]{\spacefactor3000\relax}%
\providecommand \BibitemShut  [1]{\csname bibitem#1\endcsname}%
\let\auto@bib@innerbib\@empty
%</preamble>
\bibitem [{\citenamefont {Niedenzu}\ \emph {et~al.}(2016)\citenamefont
  {Niedenzu}, \citenamefont {Gelbwaser-Klimovsky}, \citenamefont {Kofman},\
  and\ \citenamefont {Kurizki}}]{Niedenzu_2016}%
  \BibitemOpen
  \bibfield  {author} {\bibinfo {author} {\bibfnamefont {W.}~\bibnamefont
  {Niedenzu}}, \bibinfo {author} {\bibfnamefont {D.}~\bibnamefont
  {Gelbwaser-Klimovsky}}, \bibinfo {author} {\bibfnamefont {A.~G.}\
  \bibnamefont {Kofman}}, \ and\ \bibinfo {author} {\bibfnamefont
  {G.}~\bibnamefont {Kurizki}},\ }\href {\doibase
  10.1088/1367-2630/18/8/083012} {\bibfield  {journal} {\bibinfo  {journal}
  {New J. Phys.}\ }\textbf {\bibinfo {volume} {18}},\ \bibinfo {pages} {083012}
  (\bibinfo {year} {2016})}\BibitemShut {NoStop}%
\bibitem [{\citenamefont {Lostaglio}\ \emph
  {et~al.}(2015{\natexlab{a}})\citenamefont {Lostaglio}, \citenamefont
  {Jennings},\ and\ \citenamefont {Rudolph}}]{Lostaglio2015}%
  \BibitemOpen
  \bibfield  {author} {\bibinfo {author} {\bibfnamefont {M.}~\bibnamefont
  {Lostaglio}}, \bibinfo {author} {\bibfnamefont {D.}~\bibnamefont {Jennings}},
  \ and\ \bibinfo {author} {\bibfnamefont {T.}~\bibnamefont {Rudolph}},\ }\href
  {\doibase 10.1038/ncomms7383} {\bibfield  {journal} {\bibinfo  {journal}
  {Nat. Commun.}\ }\textbf {\bibinfo {volume} {6}},\ \bibinfo {pages} {6383}
  (\bibinfo {year} {2015}{\natexlab{a}})}\BibitemShut {NoStop}%
\bibitem [{\citenamefont {Lostaglio}\ \emph
  {et~al.}(2015{\natexlab{b}})\citenamefont {Lostaglio}, \citenamefont
  {Korzekwa}, \citenamefont {Jennings},\ and\ \citenamefont
  {Rudolph}}]{Lostaglio2015_2}%
  \BibitemOpen
  \bibfield  {author} {\bibinfo {author} {\bibfnamefont {M.}~\bibnamefont
  {Lostaglio}}, \bibinfo {author} {\bibfnamefont {K.}~\bibnamefont {Korzekwa}},
  \bibinfo {author} {\bibfnamefont {D.}~\bibnamefont {Jennings}}, \ and\
  \bibinfo {author} {\bibfnamefont {T.}~\bibnamefont {Rudolph}},\ }\href
  {\doibase 10.1103/PhysRevX.5.021001} {\bibfield  {journal} {\bibinfo
  {journal} {Phys. Rev. X}\ }\textbf {\bibinfo {volume} {5}},\ \bibinfo {pages}
  {021001} (\bibinfo {year} {2015}{\natexlab{b}})}\BibitemShut {NoStop}%
\bibitem [{\citenamefont {Korzekwa}\ \emph {et~al.}(2016)\citenamefont
  {Korzekwa}, \citenamefont {Lostaglio}, \citenamefont {Oppenheim},\ and\
  \citenamefont {Jennings}}]{Korzekwa_2016}%
  \BibitemOpen
  \bibfield  {author} {\bibinfo {author} {\bibfnamefont {K.}~\bibnamefont
  {Korzekwa}}, \bibinfo {author} {\bibfnamefont {M.}~\bibnamefont {Lostaglio}},
  \bibinfo {author} {\bibfnamefont {J.}~\bibnamefont {Oppenheim}}, \ and\
  \bibinfo {author} {\bibfnamefont {D.}~\bibnamefont {Jennings}},\ }\href
  {\doibase 10.1088/1367-2630/18/2/023045} {\bibfield  {journal} {\bibinfo
  {journal} {New J. Phys.}\ }\textbf {\bibinfo {volume} {18}},\ \bibinfo
  {pages} {023045} (\bibinfo {year} {2016})}\BibitemShut {NoStop}%
\bibitem [{\citenamefont {Deffner}\ and\ \citenamefont
  {Campbell}(2019)}]{10.1088/2053-2571/ab21c6}%
  \BibitemOpen
  \bibfield  {author} {\bibinfo {author} {\bibfnamefont {S.}~\bibnamefont
  {Deffner}}\ and\ \bibinfo {author} {\bibfnamefont {S.}~\bibnamefont
  {Campbell}},\ }\href {\doibase 10.1088/2053-2571/ab21c6} {\emph {\bibinfo
  {title} {Quantum Thermodynamics}}}\ (\bibinfo  {publisher} {Morgan and
  Claypool Publishers},\ \bibinfo {year} {2019})\BibitemShut {NoStop}%
\bibitem [{\citenamefont {Binder}\ \emph {et~al.}(2019)\citenamefont {Binder},
  \citenamefont {Correa}, \citenamefont {Gogolin}, \citenamefont {Anders},\
  and\ \citenamefont {Adesso}}]{binder2019thermodynamics}%
  \BibitemOpen
  \bibfield  {author} {\bibinfo {author} {\bibfnamefont {F.}~\bibnamefont
  {Binder}}, \bibinfo {author} {\bibfnamefont {L.}~\bibnamefont {Correa}},
  \bibinfo {author} {\bibfnamefont {C.}~\bibnamefont {Gogolin}}, \bibinfo
  {author} {\bibfnamefont {J.}~\bibnamefont {Anders}}, \ and\ \bibinfo {author}
  {\bibfnamefont {G.}~\bibnamefont {Adesso}},\ }\href
  {https://books.google.pl/books?id=IQlouQEACAAJ} {\emph {\bibinfo {title}
  {Thermodynamics in the Quantum Regime: Fundamental Aspects and New
  Directions}}},\ Fundamental Theories of Physics\ (\bibinfo  {publisher}
  {Springer International Publishing},\ \bibinfo {year} {2019})\BibitemShut
  {NoStop}%
\bibitem [{\citenamefont {Goold}\ \emph {et~al.}(2016)\citenamefont {Goold},
  \citenamefont {Huber}, \citenamefont {Riera}, \citenamefont {del Rio},\ and\
  \citenamefont {Skrzypczyk}}]{Goold_2016}%
  \BibitemOpen
  \bibfield  {author} {\bibinfo {author} {\bibfnamefont {J.}~\bibnamefont
  {Goold}}, \bibinfo {author} {\bibfnamefont {M.}~\bibnamefont {Huber}},
  \bibinfo {author} {\bibfnamefont {A.}~\bibnamefont {Riera}}, \bibinfo
  {author} {\bibfnamefont {L.}~\bibnamefont {del Rio}}, \ and\ \bibinfo
  {author} {\bibfnamefont {P.}~\bibnamefont {Skrzypczyk}},\ }\href {\doibase
  10.1088/1751-8113/49/14/143001} {\bibfield  {journal} {\bibinfo  {journal}
  {J. Phys. A Math. Theor.}\ }\textbf {\bibinfo {volume} {49}},\ \bibinfo
  {pages} {143001} (\bibinfo {year} {2016})}\BibitemShut {NoStop}%
\bibitem [{\citenamefont {Myers}\ \emph {et~al.}(2022)\citenamefont {Myers},
  \citenamefont {Abah},\ and\ \citenamefont {Deffner}}]{myers2022quantum}%
  \BibitemOpen
  \bibfield  {author} {\bibinfo {author} {\bibfnamefont {N.~M.}\ \bibnamefont
  {Myers}}, \bibinfo {author} {\bibfnamefont {O.}~\bibnamefont {Abah}}, \ and\
  \bibinfo {author} {\bibfnamefont {S.}~\bibnamefont {Deffner}},\ }\href
  {\doibase 10.1116/5.0083192} {\bibfield  {journal} {\bibinfo  {journal}
  {AVS}\ }\textbf {\bibinfo {volume} {4}},\ \bibinfo {pages} {027101} (\bibinfo
  {year} {2022})}\BibitemShut {NoStop}%
\bibitem [{\citenamefont {Ghosh}\ \emph {et~al.}(2019)\citenamefont {Ghosh},
  \citenamefont {Mukherjee}, \citenamefont {Niedenzu},\ and\ \citenamefont
  {Kurizki}}]{Ghosh2019}%
  \BibitemOpen
  \bibfield  {author} {\bibinfo {author} {\bibfnamefont {A.}~\bibnamefont
  {Ghosh}}, \bibinfo {author} {\bibfnamefont {V.}~\bibnamefont {Mukherjee}},
  \bibinfo {author} {\bibfnamefont {W.}~\bibnamefont {Niedenzu}}, \ and\
  \bibinfo {author} {\bibfnamefont {G.}~\bibnamefont {Kurizki}},\ }\href
  {\doibase 10.1140/epjst/e2019-800060-7} {\bibfield  {journal} {\bibinfo
  {journal} {Eur. Phys. J.: Spec. Top.}\ }\textbf {\bibinfo {volume} {227}},\
  \bibinfo {pages} {2043} (\bibinfo {year} {2019})}\BibitemShut {NoStop}%
\bibitem [{\citenamefont {Ghosh}\ \emph {et~al.}(2018)\citenamefont {Ghosh},
  \citenamefont {Gelbwaser-Klimovsky}, \citenamefont {Niedenzu}, \citenamefont
  {Lvovsky}, \citenamefont {Mazets}, \citenamefont {Scully},\ and\
  \citenamefont {Kurizki}}]{GhoshM}%
  \BibitemOpen
  \bibfield  {author} {\bibinfo {author} {\bibfnamefont {A.}~\bibnamefont
  {Ghosh}}, \bibinfo {author} {\bibfnamefont {D.}~\bibnamefont
  {Gelbwaser-Klimovsky}}, \bibinfo {author} {\bibfnamefont {W.}~\bibnamefont
  {Niedenzu}}, \bibinfo {author} {\bibfnamefont {A.~I.}\ \bibnamefont
  {Lvovsky}}, \bibinfo {author} {\bibfnamefont {I.}~\bibnamefont {Mazets}},
  \bibinfo {author} {\bibfnamefont {M.~O.}\ \bibnamefont {Scully}}, \ and\
  \bibinfo {author} {\bibfnamefont {G.}~\bibnamefont {Kurizki}},\ }\href
  {\doibase 10.1073/pnas.1805354115} {\bibfield  {journal} {\bibinfo  {journal}
  {Proc. Natl. Acad. Sci. U.S.A.}\ }\textbf {\bibinfo {volume} {115}},\
  \bibinfo {pages} {9941} (\bibinfo {year} {2018})}\BibitemShut {NoStop}%
\bibitem [{\citenamefont {Alicki}(1979)}]{Alicki}%
  \BibitemOpen
  \bibfield  {author} {\bibinfo {author} {\bibfnamefont {R.}~\bibnamefont
  {Alicki}},\ }\href {http://stacks.iop.org/0305-4470/12/i=5/a=007} {\bibfield
  {journal} {\bibinfo  {journal} {J. Phys. A Math. Theor.}\ }\textbf {\bibinfo
  {volume} {12}},\ \bibinfo {pages} {L103} (\bibinfo {year}
  {1979})}\BibitemShut {NoStop}%
\bibitem [{\citenamefont {Gardas}\ and\ \citenamefont
  {Deffner}(2015)}]{Gardas}%
  \BibitemOpen
  \bibfield  {author} {\bibinfo {author} {\bibfnamefont {B.}~\bibnamefont
  {Gardas}}\ and\ \bibinfo {author} {\bibfnamefont {S.}~\bibnamefont
  {Deffner}},\ }\href {\doibase 10.1103/PhysRevE.92.042126} {\bibfield
  {journal} {\bibinfo  {journal} {Phys. Rev. E}\ }\textbf {\bibinfo {volume}
  {92}},\ \bibinfo {pages} {042126} (\bibinfo {year} {2015})}\BibitemShut
  {NoStop}%
\bibitem [{\citenamefont {Feldmann}\ and\ \citenamefont {Kosloff}(2000)}]{Fed}%
  \BibitemOpen
  \bibfield  {author} {\bibinfo {author} {\bibfnamefont {T.}~\bibnamefont
  {Feldmann}}\ and\ \bibinfo {author} {\bibfnamefont {R.}~\bibnamefont
  {Kosloff}},\ }\href {\doibase 10.1103/PhysRevE.61.4774} {\bibfield  {journal}
  {\bibinfo  {journal} {Phys. Rev. E}\ }\textbf {\bibinfo {volume} {61}},\
  \bibinfo {pages} {4774} (\bibinfo {year} {2000})}\BibitemShut {NoStop}%
\bibitem [{\citenamefont {Correa}\ \emph {et~al.}(2014)\citenamefont {Correa},
  \citenamefont {Palao}, \citenamefont {Alonso},\ and\ \citenamefont
  {Adesso}}]{Correa2014}%
  \BibitemOpen
  \bibfield  {author} {\bibinfo {author} {\bibfnamefont {L.~A.}\ \bibnamefont
  {Correa}}, \bibinfo {author} {\bibfnamefont {J.~P.}\ \bibnamefont {Palao}},
  \bibinfo {author} {\bibfnamefont {D.}~\bibnamefont {Alonso}}, \ and\ \bibinfo
  {author} {\bibfnamefont {G.}~\bibnamefont {Adesso}},\ }\href {\doibase
  10.1038/srep03949} {\bibfield  {journal} {\bibinfo  {journal} {Sci. Rep.}\
  }\textbf {\bibinfo {volume} {4}},\ \bibinfo {pages} {3949} (\bibinfo {year}
  {2014})}\BibitemShut {NoStop}%
\bibitem [{\citenamefont {Harbola}\ \emph {et~al.}(2012)\citenamefont
  {Harbola}, \citenamefont {Rahav},\ and\ \citenamefont
  {Mukamel}}]{Harbola_2012}%
  \BibitemOpen
  \bibfield  {author} {\bibinfo {author} {\bibfnamefont {U.}~\bibnamefont
  {Harbola}}, \bibinfo {author} {\bibfnamefont {S.}~\bibnamefont {Rahav}}, \
  and\ \bibinfo {author} {\bibfnamefont {S.}~\bibnamefont {Mukamel}},\ }\href
  {\doibase 10.1209/0295-5075/99/50005} {\bibfield  {journal} {\bibinfo
  {journal} {EPL}\ }\textbf {\bibinfo {volume} {99}},\ \bibinfo {pages} {50005}
  (\bibinfo {year} {2012})}\BibitemShut {NoStop}%
\bibitem [{\citenamefont {Linden}\ \emph {et~al.}(2010)\citenamefont {Linden},
  \citenamefont {Popescu},\ and\ \citenamefont {Skrzypczyk}}]{Linden}%
  \BibitemOpen
  \bibfield  {author} {\bibinfo {author} {\bibfnamefont {N.}~\bibnamefont
  {Linden}}, \bibinfo {author} {\bibfnamefont {S.}~\bibnamefont {Popescu}}, \
  and\ \bibinfo {author} {\bibfnamefont {P.}~\bibnamefont {Skrzypczyk}},\
  }\href {\doibase 10.1103/PhysRevLett.105.130401} {\bibfield  {journal}
  {\bibinfo  {journal} {Phys. Rev. Lett.}\ }\textbf {\bibinfo {volume} {105}},\
  \bibinfo {pages} {130401} (\bibinfo {year} {2010})}\BibitemShut {NoStop}%
\bibitem [{\citenamefont {Quan}\ \emph {et~al.}(2007)\citenamefont {Quan},
  \citenamefont {Liu}, \citenamefont {Sun},\ and\ \citenamefont {Nori}}]{Quan}%
  \BibitemOpen
  \bibfield  {author} {\bibinfo {author} {\bibfnamefont {H.~T.}\ \bibnamefont
  {Quan}}, \bibinfo {author} {\bibfnamefont {Y.-x.}\ \bibnamefont {Liu}},
  \bibinfo {author} {\bibfnamefont {C.~P.}\ \bibnamefont {Sun}}, \ and\
  \bibinfo {author} {\bibfnamefont {F.}~\bibnamefont {Nori}},\ }\href {\doibase
  10.1103/PhysRevE.76.031105} {\bibfield  {journal} {\bibinfo  {journal} {Phys.
  Rev. E}\ }\textbf {\bibinfo {volume} {76}},\ \bibinfo {pages} {031105}
  (\bibinfo {year} {2007})}\BibitemShut {NoStop}%
\bibitem [{\citenamefont {Bhattacharjee}\ and\ \citenamefont
  {Dutta}(2021)}]{Sourav2020}%
  \BibitemOpen
  \bibfield  {author} {\bibinfo {author} {\bibfnamefont {S.}~\bibnamefont
  {Bhattacharjee}}\ and\ \bibinfo {author} {\bibfnamefont {A.}~\bibnamefont
  {Dutta}},\ }\href {\doibase 10.1140/epjb/s10051-021-00235-3} {\bibfield
  {journal} {\bibinfo  {journal} {Eur Phys J B}\ }\textbf {\bibinfo {volume}
  {94}},\ \bibinfo {pages} {239} (\bibinfo {year} {2021})}\BibitemShut
  {NoStop}%
\bibitem [{\citenamefont {Scully}\ \emph {et~al.}(2003)\citenamefont {Scully},
  \citenamefont {Zubairy}, \citenamefont {Agarwal},\ and\ \citenamefont
  {Walther}}]{Scully862}%
  \BibitemOpen
  \bibfield  {author} {\bibinfo {author} {\bibfnamefont {M.~O.}\ \bibnamefont
  {Scully}}, \bibinfo {author} {\bibfnamefont {M.~S.}\ \bibnamefont {Zubairy}},
  \bibinfo {author} {\bibfnamefont {G.~S.}\ \bibnamefont {Agarwal}}, \ and\
  \bibinfo {author} {\bibfnamefont {H.}~\bibnamefont {Walther}},\ }\href
  {\doibase 10.1126/science.1078955} {\bibfield  {journal} {\bibinfo  {journal}
  {Science}\ }\textbf {\bibinfo {volume} {299}},\ \bibinfo {pages} {862}
  (\bibinfo {year} {2003})}\BibitemShut {NoStop}%
\bibitem [{\citenamefont {Dillenschneider}\ and\ \citenamefont
  {Lutz}(2009)}]{Dillenschneider_2009}%
  \BibitemOpen
  \bibfield  {author} {\bibinfo {author} {\bibfnamefont {R.}~\bibnamefont
  {Dillenschneider}}\ and\ \bibinfo {author} {\bibfnamefont {E.}~\bibnamefont
  {Lutz}},\ }\href {\doibase 10.1209/0295-5075/88/50003} {\bibfield  {journal}
  {\bibinfo  {journal} {EPL.}\ }\textbf {\bibinfo {volume} {88}},\ \bibinfo
  {pages} {50003} (\bibinfo {year} {2009})}\BibitemShut {NoStop}%
\bibitem [{\citenamefont {Huang}\ \emph {et~al.}(2012)\citenamefont {Huang},
  \citenamefont {Wang},\ and\ \citenamefont {Yi}}]{PhysRevE.86.051105}%
  \BibitemOpen
  \bibfield  {author} {\bibinfo {author} {\bibfnamefont {X.~L.}\ \bibnamefont
  {Huang}}, \bibinfo {author} {\bibfnamefont {T.}~\bibnamefont {Wang}}, \ and\
  \bibinfo {author} {\bibfnamefont {X.~X.}\ \bibnamefont {Yi}},\ }\href
  {\doibase 10.1103/PhysRevE.86.051105} {\bibfield  {journal} {\bibinfo
  {journal} {Phys. Rev. E}\ }\textbf {\bibinfo {volume} {86}},\ \bibinfo
  {pages} {051105} (\bibinfo {year} {2012})}\BibitemShut {NoStop}%
\bibitem [{\citenamefont {Abah}\ and\ \citenamefont {Lutz}(2014)}]{Abah_2014}%
  \BibitemOpen
  \bibfield  {author} {\bibinfo {author} {\bibfnamefont {O.}~\bibnamefont
  {Abah}}\ and\ \bibinfo {author} {\bibfnamefont {E.}~\bibnamefont {Lutz}},\
  }\href {\doibase 10.1209/0295-5075/106/20001} {\bibfield  {journal} {\bibinfo
   {journal} {EPL}\ }\textbf {\bibinfo {volume} {106}},\ \bibinfo {pages}
  {20001} (\bibinfo {year} {2014})}\BibitemShut {NoStop}%
\bibitem [{\citenamefont {Niedenzu}\ \emph {et~al.}(2018)\citenamefont
  {Niedenzu}, \citenamefont {Mukherjee}, \citenamefont {Ghosh}, \citenamefont
  {Kofman},\ and\ \citenamefont {Kurizki}}]{Niedenzu2018}%
  \BibitemOpen
  \bibfield  {author} {\bibinfo {author} {\bibfnamefont {W.}~\bibnamefont
  {Niedenzu}}, \bibinfo {author} {\bibfnamefont {V.}~\bibnamefont {Mukherjee}},
  \bibinfo {author} {\bibfnamefont {A.}~\bibnamefont {Ghosh}}, \bibinfo
  {author} {\bibfnamefont {A.~G.}\ \bibnamefont {Kofman}}, \ and\ \bibinfo
  {author} {\bibfnamefont {G.}~\bibnamefont {Kurizki}},\ }\href {\doibase
  10.1038/s41467-017-01991-6} {\bibfield  {journal} {\bibinfo  {journal} {Nat.
  Commun.}\ }\textbf {\bibinfo {volume} {9}},\ \bibinfo {pages} {165} (\bibinfo
  {year} {2018})}\BibitemShut {NoStop}%
\bibitem [{\citenamefont {Manzano}\ \emph {et~al.}(2016)\citenamefont
  {Manzano}, \citenamefont {Galve}, \citenamefont {Zambrini},\ and\
  \citenamefont {Parrondo}}]{Gonzalo1}%
  \BibitemOpen
  \bibfield  {author} {\bibinfo {author} {\bibfnamefont {G.}~\bibnamefont
  {Manzano}}, \bibinfo {author} {\bibfnamefont {F.}~\bibnamefont {Galve}},
  \bibinfo {author} {\bibfnamefont {R.}~\bibnamefont {Zambrini}}, \ and\
  \bibinfo {author} {\bibfnamefont {J.~M.~R.}\ \bibnamefont {Parrondo}},\
  }\href {\doibase 10.1103/PhysRevE.93.052120} {\bibfield  {journal} {\bibinfo
  {journal} {Phys. Rev. E}\ }\textbf {\bibinfo {volume} {93}},\ \bibinfo
  {pages} {052120} (\bibinfo {year} {2016})}\BibitemShut {NoStop}%
\bibitem [{\citenamefont {Wang}\ \emph {et~al.}(2019)\citenamefont {Wang},
  \citenamefont {He},\ and\ \citenamefont {Ma}}]{Wang2019}%
  \BibitemOpen
  \bibfield  {author} {\bibinfo {author} {\bibfnamefont {J.}~\bibnamefont
  {Wang}}, \bibinfo {author} {\bibfnamefont {J.}~\bibnamefont {He}}, \ and\
  \bibinfo {author} {\bibfnamefont {Y.}~\bibnamefont {Ma}},\ }\href {\doibase
  10.1103/PhysRevE.100.052126} {\bibfield  {journal} {\bibinfo  {journal}
  {Phys. Rev. E}\ }\textbf {\bibinfo {volume} {100}},\ \bibinfo {pages}
  {052126} (\bibinfo {year} {2019})}\BibitemShut {NoStop}%
\bibitem [{\citenamefont {Singh}\ and\ \citenamefont {M\"ustecapl\ifmmode
  \imath \else \i \fi{}o\ifmmode~\breve{g}\else
  \u{g}\fi{}lu}(2020)}]{Singh2020}%
  \BibitemOpen
  \bibfield  {author} {\bibinfo {author} {\bibfnamefont {V.}~\bibnamefont
  {Singh}}\ and\ \bibinfo {author} {\bibfnamefont {O.~E.}\ \bibnamefont
  {M\"ustecapl\ifmmode \imath \else \i \fi{}o\ifmmode~\breve{g}\else
  \u{g}\fi{}lu}},\ }\href {\doibase 10.1103/PhysRevE.102.062123} {\bibfield
  {journal} {\bibinfo  {journal} {Phys. Rev. E}\ }\textbf {\bibinfo {volume}
  {102}},\ \bibinfo {pages} {062123} (\bibinfo {year} {2020})}\BibitemShut
  {NoStop}%
\bibitem [{\citenamefont {Xiao}\ and\ \citenamefont {Li}(2018)}]{XIAO20183051}%
  \BibitemOpen
  \bibfield  {author} {\bibinfo {author} {\bibfnamefont {B.}~\bibnamefont
  {Xiao}}\ and\ \bibinfo {author} {\bibfnamefont {R.}~\bibnamefont {Li}},\
  }\href {\doibase https://doi.org/10.1016/j.physleta.2018.07.033} {\bibfield
  {journal} {\bibinfo  {journal} {Phys. Lett. A}\ }\textbf {\bibinfo {volume}
  {382}},\ \bibinfo {pages} {3051 } (\bibinfo {year} {2018})}\BibitemShut
  {NoStop}%
\bibitem [{\citenamefont {Ro\ss{}nagel}\ \emph {et~al.}(2014)\citenamefont
  {Ro\ss{}nagel}, \citenamefont {Abah}, \citenamefont {Schmidt-Kaler},
  \citenamefont {Singer},\ and\ \citenamefont {Lutz}}]{Abah}%
  \BibitemOpen
  \bibfield  {author} {\bibinfo {author} {\bibfnamefont {J.}~\bibnamefont
  {Ro\ss{}nagel}}, \bibinfo {author} {\bibfnamefont {O.}~\bibnamefont {Abah}},
  \bibinfo {author} {\bibfnamefont {F.}~\bibnamefont {Schmidt-Kaler}}, \bibinfo
  {author} {\bibfnamefont {K.}~\bibnamefont {Singer}}, \ and\ \bibinfo {author}
  {\bibfnamefont {E.}~\bibnamefont {Lutz}},\ }\href {\doibase
  10.1103/PhysRevLett.112.030602} {\bibfield  {journal} {\bibinfo  {journal}
  {Phys. Rev. Lett.}\ }\textbf {\bibinfo {volume} {112}},\ \bibinfo {pages}
  {030602} (\bibinfo {year} {2014})}\BibitemShut {NoStop}%
\bibitem [{\citenamefont {Klaers}\ \emph {et~al.}(2017)\citenamefont {Klaers},
  \citenamefont {Faelt}, \citenamefont {Imamoglu},\ and\ \citenamefont
  {Togan}}]{PhysRevX.7.031044}%
  \BibitemOpen
  \bibfield  {author} {\bibinfo {author} {\bibfnamefont {J.}~\bibnamefont
  {Klaers}}, \bibinfo {author} {\bibfnamefont {S.}~\bibnamefont {Faelt}},
  \bibinfo {author} {\bibfnamefont {A.}~\bibnamefont {Imamoglu}}, \ and\
  \bibinfo {author} {\bibfnamefont {E.}~\bibnamefont {Togan}},\ }\href
  {\doibase 10.1103/PhysRevX.7.031044} {\bibfield  {journal} {\bibinfo
  {journal} {Phys. Rev. X}\ }\textbf {\bibinfo {volume} {7}},\ \bibinfo {pages}
  {031044} (\bibinfo {year} {2017})}\BibitemShut {NoStop}%
\bibitem [{\citenamefont {Manzano}(2018)}]{Gonzalo}%
  \BibitemOpen
  \bibfield  {author} {\bibinfo {author} {\bibfnamefont {G.}~\bibnamefont
  {Manzano}},\ }\href {\doibase 10.1103/PhysRevE.98.042123} {\bibfield
  {journal} {\bibinfo  {journal} {Phys. Rev. E}\ }\textbf {\bibinfo {volume}
  {98}},\ \bibinfo {pages} {042123} (\bibinfo {year} {2018})}\BibitemShut
  {NoStop}%
\bibitem [{\citenamefont {Pusz}\ and\ \citenamefont
  {Woronowicz}(1978)}]{Pusz1978}%
  \BibitemOpen
  \bibfield  {author} {\bibinfo {author} {\bibfnamefont {W.}~\bibnamefont
  {Pusz}}\ and\ \bibinfo {author} {\bibfnamefont {S.~L.}\ \bibnamefont
  {Woronowicz}},\ }\href {\doibase 10.1007/BF01614224} {\bibfield  {journal}
  {\bibinfo  {journal} {Commun. Math. Phys.}\ }\textbf {\bibinfo {volume}
  {58}},\ \bibinfo {pages} {273} (\bibinfo {year} {1978})}\BibitemShut
  {NoStop}%
\bibitem [{\citenamefont {Lenard}(1978)}]{Lenard1978}%
  \BibitemOpen
  \bibfield  {author} {\bibinfo {author} {\bibfnamefont {A.}~\bibnamefont
  {Lenard}},\ }\href {\doibase 10.1007/BF01011769} {\bibfield  {journal}
  {\bibinfo  {journal} {J. Stat. Phys.}\ }\textbf {\bibinfo {volume} {19}},\
  \bibinfo {pages} {575} (\bibinfo {year} {1978})}\BibitemShut {NoStop}%
\bibitem [{\citenamefont {Dodonov}\ and\ \citenamefont
  {Man'ko}(2003)}]{dodonov2003theory}%
  \BibitemOpen
  \bibfield  {author} {\bibinfo {author} {\bibfnamefont {V.}~\bibnamefont
  {Dodonov}}\ and\ \bibinfo {author} {\bibfnamefont {V.}~\bibnamefont
  {Man'ko}},\ }\href {https://books.google.pl/books?id=XIiBgb0dfJwC} {\emph
  {\bibinfo {title} {Theory of Nonclassical States of Light}}}\ (\bibinfo
  {publisher} {Taylor \& Francis},\ \bibinfo {year} {2003})\BibitemShut
  {NoStop}%
\bibitem [{\citenamefont {Lee}(1991)}]{Ctlee}%
  \BibitemOpen
  \bibfield  {author} {\bibinfo {author} {\bibfnamefont {C.~T.}\ \bibnamefont
  {Lee}},\ }\href {\doibase 10.1103/PhysRevA.44.R2775} {\bibfield  {journal}
  {\bibinfo  {journal} {Phys. Rev. A}\ }\textbf {\bibinfo {volume} {44}},\
  \bibinfo {pages} {R2775} (\bibinfo {year} {1991})}\BibitemShut {NoStop}%
\bibitem [{\citenamefont {de~Oliveira}(2004)}]{Marcos1}%
  \BibitemOpen
  \bibfield  {author} {\bibinfo {author} {\bibfnamefont {M.~C.}\ \bibnamefont
  {de~Oliveira}},\ }\href {\doibase 10.1103/PhysRevA.70.034303} {\bibfield
  {journal} {\bibinfo  {journal} {Phys. Rev. A}\ }\textbf {\bibinfo {volume}
  {70}},\ \bibinfo {pages} {034303} (\bibinfo {year} {2004})}\BibitemShut
  {NoStop}%
\bibitem [{\citenamefont {Kim}\ \emph {et~al.}(2002)\citenamefont {Kim},
  \citenamefont {Son}, \citenamefont {Bu\ifmmode~\check{z}\else \v{z}\fi{}ek},\
  and\ \citenamefont {Knight}}]{Kim}%
  \BibitemOpen
  \bibfield  {author} {\bibinfo {author} {\bibfnamefont {M.~S.}\ \bibnamefont
  {Kim}}, \bibinfo {author} {\bibfnamefont {W.}~\bibnamefont {Son}}, \bibinfo
  {author} {\bibfnamefont {V.}~\bibnamefont {Bu\ifmmode~\check{z}\else
  \v{z}\fi{}ek}}, \ and\ \bibinfo {author} {\bibfnamefont {P.~L.}\ \bibnamefont
  {Knight}},\ }\href {\doibase 10.1103/PhysRevA.65.032323} {\bibfield
  {journal} {\bibinfo  {journal} {Phys. Rev. A}\ }\textbf {\bibinfo {volume}
  {65}},\ \bibinfo {pages} {032323} (\bibinfo {year} {2002})}\BibitemShut
  {NoStop}%
\bibitem [{\citenamefont {Xiang-bin}(2002)}]{Xiang-bin}%
  \BibitemOpen
  \bibfield  {author} {\bibinfo {author} {\bibfnamefont {W.}~\bibnamefont
  {Xiang-bin}},\ }\href {\doibase 10.1103/PhysRevA.66.024303} {\bibfield
  {journal} {\bibinfo  {journal} {Phys. Rev. A}\ }\textbf {\bibinfo {volume}
  {66}},\ \bibinfo {pages} {024303} (\bibinfo {year} {2002})}\BibitemShut
  {NoStop}%
\bibitem [{\citenamefont {{de Oliveira}}\ and\ \citenamefont
  {Munro}(2004)}]{Nclass}%
  \BibitemOpen
  \bibfield  {author} {\bibinfo {author} {\bibfnamefont {M.}~\bibnamefont {{de
  Oliveira}}}\ and\ \bibinfo {author} {\bibfnamefont {W.}~\bibnamefont
  {Munro}},\ }\href {\doibase https://doi.org/10.1016/j.physleta.2003.11.037}
  {\bibfield  {journal} {\bibinfo  {journal} {Phys. Lett. A}\ }\textbf
  {\bibinfo {volume} {320}},\ \bibinfo {pages} {352 } (\bibinfo {year}
  {2004})}\BibitemShut {NoStop}%
\bibitem [{\citenamefont {Scully}\ and\ \citenamefont
  {Zubairy}(1997)}]{scully1997quantum}%
  \BibitemOpen
  \bibfield  {author} {\bibinfo {author} {\bibfnamefont {M.}~\bibnamefont
  {Scully}}\ and\ \bibinfo {author} {\bibfnamefont {M.}~\bibnamefont
  {Zubairy}},\ }\href {https://books.google.com.br/books?id=20ISsQCKKmQC}
  {\emph {\bibinfo {title} {Quantum Optics}}}\ (\bibinfo  {publisher}
  {Cambridge University Press},\ \bibinfo {year} {1997})\BibitemShut {NoStop}%
\bibitem [{\citenamefont {Slusher}\ \emph {et~al.}(1985)\citenamefont
  {Slusher}, \citenamefont {Hollberg}, \citenamefont {Yurke}, \citenamefont
  {Mertz},\ and\ \citenamefont {Valley}}]{nonl1}%
  \BibitemOpen
  \bibfield  {author} {\bibinfo {author} {\bibfnamefont {R.~E.}\ \bibnamefont
  {Slusher}}, \bibinfo {author} {\bibfnamefont {L.~W.}\ \bibnamefont
  {Hollberg}}, \bibinfo {author} {\bibfnamefont {B.}~\bibnamefont {Yurke}},
  \bibinfo {author} {\bibfnamefont {J.~C.}\ \bibnamefont {Mertz}}, \ and\
  \bibinfo {author} {\bibfnamefont {J.~F.}\ \bibnamefont {Valley}},\ }\href
  {\doibase 10.1103/PhysRevLett.55.2409} {\bibfield  {journal} {\bibinfo
  {journal} {Phys. Rev. Lett.}\ }\textbf {\bibinfo {volume} {55}},\ \bibinfo
  {pages} {2409} (\bibinfo {year} {1985})}\BibitemShut {NoStop}%
\bibitem [{\citenamefont {Shelby}\ \emph {et~al.}(1986)\citenamefont {Shelby},
  \citenamefont {Levenson}, \citenamefont {Perlmutter}, \citenamefont {DeVoe},\
  and\ \citenamefont {Walls}}]{nonl2}%
  \BibitemOpen
  \bibfield  {author} {\bibinfo {author} {\bibfnamefont {R.~M.}\ \bibnamefont
  {Shelby}}, \bibinfo {author} {\bibfnamefont {M.~D.}\ \bibnamefont
  {Levenson}}, \bibinfo {author} {\bibfnamefont {S.~H.}\ \bibnamefont
  {Perlmutter}}, \bibinfo {author} {\bibfnamefont {R.~G.}\ \bibnamefont
  {DeVoe}}, \ and\ \bibinfo {author} {\bibfnamefont {D.~F.}\ \bibnamefont
  {Walls}},\ }\href {\doibase 10.1103/PhysRevLett.57.691} {\bibfield  {journal}
  {\bibinfo  {journal} {Phys. Rev. Lett.}\ }\textbf {\bibinfo {volume} {57}},\
  \bibinfo {pages} {691} (\bibinfo {year} {1986})}\BibitemShut {NoStop}%
\bibitem [{\citenamefont {Wu}\ \emph {et~al.}(1986)\citenamefont {Wu},
  \citenamefont {Kimble}, \citenamefont {Hall},\ and\ \citenamefont
  {Wu}}]{nonl3}%
  \BibitemOpen
  \bibfield  {author} {\bibinfo {author} {\bibfnamefont {L.-A.}\ \bibnamefont
  {Wu}}, \bibinfo {author} {\bibfnamefont {H.~J.}\ \bibnamefont {Kimble}},
  \bibinfo {author} {\bibfnamefont {J.~L.}\ \bibnamefont {Hall}}, \ and\
  \bibinfo {author} {\bibfnamefont {H.}~\bibnamefont {Wu}},\ }\href {\doibase
  10.1103/PhysRevLett.57.2520} {\bibfield  {journal} {\bibinfo  {journal}
  {Phys. Rev. Lett.}\ }\textbf {\bibinfo {volume} {57}},\ \bibinfo {pages}
  {2520} (\bibinfo {year} {1986})}\BibitemShut {NoStop}%
\bibitem [{\citenamefont {Serafini}(2017)}]{AlessioB}%
  \BibitemOpen
  \bibfield  {author} {\bibinfo {author} {\bibfnamefont {A.}~\bibnamefont
  {Serafini}},\ }\href {https://books.google.com.br/books?id=zHtgvgAACAAJ}
  {\emph {\bibinfo {title} {Quantum Continuous Variables: A Primer of
  Theoretical Methods}}}\ (\bibinfo  {publisher} {CRC Press, Taylor \& Francis
  Group},\ \bibinfo {year} {2017})\BibitemShut {NoStop}%
\bibitem [{\citenamefont {Singh}\ \emph {et~al.}(2019)\citenamefont {Singh},
  \citenamefont {Jabbour}, \citenamefont {Van~Herstraeten},\ and\ \citenamefont
  {Cerf}}]{Cerf}%
  \BibitemOpen
  \bibfield  {author} {\bibinfo {author} {\bibfnamefont {U.}~\bibnamefont
  {Singh}}, \bibinfo {author} {\bibfnamefont {M.~G.}\ \bibnamefont {Jabbour}},
  \bibinfo {author} {\bibfnamefont {Z.}~\bibnamefont {Van~Herstraeten}}, \ and\
  \bibinfo {author} {\bibfnamefont {N.~J.}\ \bibnamefont {Cerf}},\ }\href
  {\doibase 10.1103/PhysRevA.100.042104} {\bibfield  {journal} {\bibinfo
  {journal} {Phys. Rev. A}\ }\textbf {\bibinfo {volume} {100}},\ \bibinfo
  {pages} {042104} (\bibinfo {year} {2019})}\BibitemShut {NoStop}%
\bibitem [{\citenamefont {Englert}\ and\ \citenamefont
  {Wódkiewicz}(2003)}]{Englert}%
  \BibitemOpen
  \bibfield  {author} {\bibinfo {author} {\bibfnamefont {B.-G.}\ \bibnamefont
  {Englert}}\ and\ \bibinfo {author} {\bibfnamefont {K.}~\bibnamefont
  {Wódkiewicz}},\ }\href {\doibase 10.1142/S0219749903000206} {\bibfield
  {journal} {\bibinfo  {journal} {Int. J. Quantum Inf.}\ }\textbf {\bibinfo
  {volume} {01}},\ \bibinfo {pages} {153} (\bibinfo {year} {2003})}\BibitemShut
  {NoStop}%
\bibitem [{\citenamefont {Koukoulekidis}\ \emph {et~al.}(2021)\citenamefont
  {Koukoulekidis}, \citenamefont {Alexander}, \citenamefont {Hebdige},\ and\
  \citenamefont {Jennings}}]{koukoulekidis2021geometry}%
  \BibitemOpen
  \bibfield  {author} {\bibinfo {author} {\bibfnamefont {N.}~\bibnamefont
  {Koukoulekidis}}, \bibinfo {author} {\bibfnamefont {R.}~\bibnamefont
  {Alexander}}, \bibinfo {author} {\bibfnamefont {T.}~\bibnamefont {Hebdige}},
  \ and\ \bibinfo {author} {\bibfnamefont {D.}~\bibnamefont {Jennings}},\
  }\href {\doibase https://doi.org/10.22331/q-2021-03-15-411} {\bibfield
  {journal} {\bibinfo  {journal} {Quantum}\ }\textbf {\bibinfo {volume} {5}},\
  \bibinfo {pages} {411} (\bibinfo {year} {2021})}\BibitemShut {NoStop}%
\bibitem [{\citenamefont {Brown}\ \emph {et~al.}(2016)\citenamefont {Brown},
  \citenamefont {Friis},\ and\ \citenamefont {Huber}}]{brown2016passivity}%
  \BibitemOpen
  \bibfield  {author} {\bibinfo {author} {\bibfnamefont {E.~G.}\ \bibnamefont
  {Brown}}, \bibinfo {author} {\bibfnamefont {N.}~\bibnamefont {Friis}}, \ and\
  \bibinfo {author} {\bibfnamefont {M.}~\bibnamefont {Huber}},\ }\href
  {https://iopscience.iop.org/article/10.1088/1367-2630/18/11/113028/meta}
  {\bibfield  {journal} {\bibinfo  {journal} {New J. Phys.}\ }\textbf {\bibinfo
  {volume} {18}},\ \bibinfo {pages} {113028} (\bibinfo {year}
  {2016})}\BibitemShut {NoStop}%
\bibitem [{\citenamefont {Breuer}\ \emph {et~al.}(2002)\citenamefont {Breuer},
  \citenamefont {Breuer}, \citenamefont {Petruccione},\ and\ \citenamefont
  {Petruccione}}]{breuer2002theory}%
  \BibitemOpen
  \bibfield  {author} {\bibinfo {author} {\bibfnamefont {H.}~\bibnamefont
  {Breuer}}, \bibinfo {author} {\bibfnamefont {P.}~\bibnamefont {Breuer}},
  \bibinfo {author} {\bibfnamefont {F.}~\bibnamefont {Petruccione}}, \ and\
  \bibinfo {author} {\bibfnamefont {S.}~\bibnamefont {Petruccione}},\ }\href
  {https://books.google.pl/books?id=0Yx5VzaMYm8C} {\emph {\bibinfo {title} {The
  Theory of Open Quantum Systems}}}\ (\bibinfo  {publisher} {Oxford University
  Press},\ \bibinfo {year} {2002})\BibitemShut {NoStop}%
\bibitem [{\citenamefont {Talkner}\ and\ \citenamefont
  {H{\"a}nggi}(2016)}]{talkner2016aspects}%
  \BibitemOpen
  \bibfield  {author} {\bibinfo {author} {\bibfnamefont {P.}~\bibnamefont
  {Talkner}}\ and\ \bibinfo {author} {\bibfnamefont {P.}~\bibnamefont
  {H{\"a}nggi}},\ }\href@noop {} {\bibfield  {journal} {\bibinfo  {journal}
  {Physical Review E}\ }\textbf {\bibinfo {volume} {93}},\ \bibinfo {pages}
  {022131} (\bibinfo {year} {2016})}\BibitemShut {NoStop}%
\bibitem [{\citenamefont {Jarzynski}(2007)}]{JARZYNSKI2007495}%
  \BibitemOpen
  \bibfield  {author} {\bibinfo {author} {\bibfnamefont {C.}~\bibnamefont
  {Jarzynski}},\ }\href {\doibase https://doi.org/10.1016/j.crhy.2007.04.010}
  {\bibfield  {journal} {\bibinfo  {journal} {C R Phys .}\ }\textbf {\bibinfo
  {volume} {8}},\ \bibinfo {pages} {495} (\bibinfo {year} {2007})}\BibitemShut
  {NoStop}%
\bibitem [{\citenamefont {Cuzminschi}\ \emph {et~al.}(2021)\citenamefont
  {Cuzminschi}, \citenamefont {Zubarev},\ and\ \citenamefont
  {Isar}}]{cuzminschi2021extractable}%
  \BibitemOpen
  \bibfield  {author} {\bibinfo {author} {\bibfnamefont {M.}~\bibnamefont
  {Cuzminschi}}, \bibinfo {author} {\bibfnamefont {A.}~\bibnamefont {Zubarev}},
  \ and\ \bibinfo {author} {\bibfnamefont {A.}~\bibnamefont {Isar}},\ }\href
  {\doibase https://doi.org/10.1038/s41598-021-03752-4} {\bibfield  {journal}
  {\bibinfo  {journal} {Sci. Rep.}\ }\textbf {\bibinfo {volume} {11}},\
  \bibinfo {pages} {1} (\bibinfo {year} {2021})}\BibitemShut {NoStop}%
\bibitem [{\citenamefont {{Horodecki}}\ and\ \citenamefont
  {{Oppenheim}}(2013)}]{horodecki2013fundamental}%
  \BibitemOpen
  \bibfield  {author} {\bibinfo {author} {\bibfnamefont {M.}~\bibnamefont
  {{Horodecki}}}\ and\ \bibinfo {author} {\bibfnamefont {J.}~\bibnamefont
  {{Oppenheim}}},\ }\href {https://www.nature.com/articles/ncomms3059}
  {\bibfield  {journal} {\bibinfo  {journal} {Nat. Commun.}\ }\textbf {\bibinfo
  {volume} {4}},\ \bibinfo {eid} {2059} (\bibinfo {year} {2013})}\BibitemShut
  {NoStop}%
\bibitem [{\citenamefont {Brand\~ao}\ \emph {et~al.}(2013)\citenamefont
  {Brand\~ao}, \citenamefont {Horodecki}, \citenamefont {Oppenheim},
  \citenamefont {Renes},\ and\ \citenamefont {Spekkens}}]{brandao2013resource}%
  \BibitemOpen
  \bibfield  {author} {\bibinfo {author} {\bibfnamefont {F.~G. S.~L.}\
  \bibnamefont {Brand\~ao}}, \bibinfo {author} {\bibfnamefont {M.}~\bibnamefont
  {Horodecki}}, \bibinfo {author} {\bibfnamefont {J.}~\bibnamefont
  {Oppenheim}}, \bibinfo {author} {\bibfnamefont {J.~M.}\ \bibnamefont
  {Renes}}, \ and\ \bibinfo {author} {\bibfnamefont {R.~W.}\ \bibnamefont
  {Spekkens}},\ }\href {https://doi.org/10.1103/PhysRevLett.111.250404}
  {\bibfield  {journal} {\bibinfo  {journal} {Phys. Rev. Lett.}\ }\textbf
  {\bibinfo {volume} {111}},\ \bibinfo {pages} {250404} (\bibinfo {year}
  {2013})}\BibitemShut {NoStop}%
\bibitem [{\citenamefont {Baumgratz}\ \emph {et~al.}(2014)\citenamefont
  {Baumgratz}, \citenamefont {Cramer},\ and\ \citenamefont {Plenio}}]{Plenio}%
  \BibitemOpen
  \bibfield  {author} {\bibinfo {author} {\bibfnamefont {T.}~\bibnamefont
  {Baumgratz}}, \bibinfo {author} {\bibfnamefont {M.}~\bibnamefont {Cramer}}, \
  and\ \bibinfo {author} {\bibfnamefont {M.~B.}\ \bibnamefont {Plenio}},\
  }\href {\doibase 10.1103/PhysRevLett.113.140401} {\bibfield  {journal}
  {\bibinfo  {journal} {Phys. Rev. Lett.}\ }\textbf {\bibinfo {volume} {113}},\
  \bibinfo {pages} {140401} (\bibinfo {year} {2014})}\BibitemShut {NoStop}%
\bibitem [{\citenamefont {Esposito}\ \emph {et~al.}(2009)\citenamefont
  {Esposito}, \citenamefont {Harbola},\ and\ \citenamefont
  {Mukamel}}]{Esposito}%
  \BibitemOpen
  \bibfield  {author} {\bibinfo {author} {\bibfnamefont {M.}~\bibnamefont
  {Esposito}}, \bibinfo {author} {\bibfnamefont {U.}~\bibnamefont {Harbola}}, \
  and\ \bibinfo {author} {\bibfnamefont {S.}~\bibnamefont {Mukamel}},\ }\href
  {\doibase 10.1103/RevModPhys.81.1665} {\bibfield  {journal} {\bibinfo
  {journal} {Rev. Mod. Phys.}\ }\textbf {\bibinfo {volume} {81}},\ \bibinfo
  {pages} {1665} (\bibinfo {year} {2009})}\BibitemShut {NoStop}%
\bibitem [{\citenamefont {Parrondo}\ \emph {et~al.}(2015)\citenamefont
  {Parrondo}, \citenamefont {Horowitz},\ and\ \citenamefont
  {Sagawa}}]{Parrondo2015}%
  \BibitemOpen
  \bibfield  {author} {\bibinfo {author} {\bibfnamefont {J.~M.~R.}\
  \bibnamefont {Parrondo}}, \bibinfo {author} {\bibfnamefont {J.~M.}\
  \bibnamefont {Horowitz}}, \ and\ \bibinfo {author} {\bibfnamefont
  {T.}~\bibnamefont {Sagawa}},\ }\href {\doibase 10.1038/nphys3230} {\bibfield
  {journal} {\bibinfo  {journal} {Nat. Phys.}\ }\textbf {\bibinfo {volume}
  {11}},\ \bibinfo {pages} {131} (\bibinfo {year} {2015})}\BibitemShut
  {NoStop}%
\bibitem [{\citenamefont {Gerry}\ \emph {et~al.}(2005)\citenamefont {Gerry},
  \citenamefont {Knight},\ and\ \citenamefont
  {Knight}}]{gerry2005introductory}%
  \BibitemOpen
  \bibfield  {author} {\bibinfo {author} {\bibfnamefont {C.}~\bibnamefont
  {Gerry}}, \bibinfo {author} {\bibfnamefont {P.}~\bibnamefont {Knight}}, \
  and\ \bibinfo {author} {\bibfnamefont {P.}~\bibnamefont {Knight}},\ }\href
  {https://books.google.com.br/books?id=CgByyoBJJwgC} {\emph {\bibinfo {title}
  {Introductory Quantum Optics}}}\ (\bibinfo  {publisher} {Cambridge University
  Press},\ \bibinfo {year} {2005})\BibitemShut {NoStop}%
\bibitem [{\citenamefont {Glauber}(1963)}]{Glauber1963}%
  \BibitemOpen
  \bibfield  {author} {\bibinfo {author} {\bibfnamefont {R.~J.}\ \bibnamefont
  {Glauber}},\ }\href {\doibase 10.1103/PhysRev.131.2766} {\bibfield  {journal}
  {\bibinfo  {journal} {Phys. Rev.}\ }\textbf {\bibinfo {volume} {131}},\
  \bibinfo {pages} {2766} (\bibinfo {year} {1963})}\BibitemShut {NoStop}%
\bibitem [{\citenamefont {Sudarshan}(1963)}]{Sudarshan:1963ts}%
  \BibitemOpen
  \bibfield  {author} {\bibinfo {author} {\bibfnamefont {E.~C.~G.}\
  \bibnamefont {Sudarshan}},\ }\href {\doibase 10.1103/PhysRevLett.10.277}
  {\bibfield  {journal} {\bibinfo  {journal} {Phys. Rev. Lett.}\ }\textbf
  {\bibinfo {volume} {10}},\ \bibinfo {pages} {277} (\bibinfo {year}
  {1963})}\BibitemShut {NoStop}%
%%CITATION = PRLTA,10,277;%%
\end{thebibliography}%
\appendix

\section{P-representable positive Gaussian operators}\label{appendix}
In this section, we derive Eq.~\eqref{Condition} from the main text. This equation gives the necessary and sufficient condition for the $P$-representability of a single bosonic mode. 

We begin by recalling that the characteristic function of a random variable ultimately determines its probability distribution. If a random variable admits a probability density function, then the characteristic function is its Fourier transform. The characteristic function $\chi_W$ of a Gaussian state can be recast in terms of its covariance matrix $\textbf{V}$ as~\cite{scully1997quantum,gerry2005introductory, Englert}
\begin{equation}
\label{CFG}
\chi_W(\z) = e^{-\frac{1}{2}\zd \textbf{V} \z} ,
\end{equation} 
where $\textbf{a} = (\alpha, \alpha^*)^{\T}$, and $\alpha$ is the complex eigenvalue of the annihilation operator $a$, whose eigenstate is the coherent state $\ket{a}$.  

A positive Gaussian operator is said to be $P$-representable if it can be written as a mixture of coherent states with a proper probability distribution function. Otherwise, the state is not $P$-representable and, therefore, non-classical~\cite{Glauber1963,Sudarshan:1963ts}. The discrimination on whether a given state is classical or not is not an easy task. However, for Gaussian states, a specific criteria can be derived~\cite{Englert}. For this purpose, let us rewrite the $P$-function in terms of the parametrisation that we have been discussing,
\begin{equation}
\label{Pfunctionp}
P (\z) = \int d\z' \: e^{-\zd \sigma_z \z'} \chi(\z')   \, ,
\end{equation}
where $\sigma_z$ is the $z$-Pauli matrix, and $\chi_{N}(\z)$ is the normally-ordered characteristic function that can be rewritten as
\begin{equation}
\label{characteristicNC}
\chi_N(\z) = e^{\frac{1}{4}\zd \z} \, \chi_W(\z) = e^{\frac{1}{4}\zd \z} \, e^{-\frac{1}{2}\zd \textbf{V} \z} = e^{\frac{1}{2}\zd \left(\frac{1}{2}\mathbbm{1}-\textbf{V} \right)\z} \, .   
\end{equation}
Substituting Eq.(\ref{characteristicNC}) into (\ref{Pfunctionp}), and using the fact that $-\zd \sigma_z \z' = -\zd{'} \sigma_z \z$~\cite{Englert}, the P-function can be written as 
\begin{equation}
\label{Pfunction1}
P(\z) = \int d\z' \: e^{\zd{'} \left( \textbf{V}- \frac{\mathbbm{1}}{2}\right)\z'} = \frac{1}{\sqrt{\textrm{det}\left(\textbf{V}-\frac{\mathbbm{1}}{2} \right) }}e^{\frac{1}{2}(\sigma_z \z)^{\T} \sigma_x \left(\textbf{V}-\frac{\mathbbm{1}}{2} \right)^{-1} (\sigma_z\z)  } \, ,
\end{equation} 
where $\sigma_x$ is the $x$-Pauli matrix. It can be verified that $(\sigma_z \z)^{\T} = \z \sigma_z$ and $\z \sigma_z \textbf{T} = -\zd \sigma_z$, thus
\begin{equation}
P(\z) = \frac{1}{\sqrt{\textrm{det}\left(\textbf{V}-\frac{\mathbbm{1}}{2} \right) }}e^{-\frac{1}{2}\zd\sigma_z\left(\textbf{V}-\frac{\mathbbm{1}}{2} \right)\sigma_z \z } \equiv \sqrt{\textrm{det}( \textbf{P})}\,e^{-\frac{1}{2}\zd \textbf{P}\z} \, ,
\end{equation}
where
\begin{equation}
\textbf{P} = \sigma_z \left(\textbf{V}-\frac{\mathbbm{1}}{2} \right)^{-1} \sigma_z \, ,
\end{equation}

From this result we may conclude that, for a single-mode Gaussian state described by a covariance matrix $\textbf{V}$ is said to be $P$-representable if
\begin{equation}
\bar{n} >  |m| \, ,
\end{equation}

Note that a $P$-representable state, $\textbf{P}$ must be non-negative, which means that $(\textbf{V}-\mathbbm{1}/2) \geq 0$. This condition requires eigenvalues greater than or equal to zero, or alternatively, a non-negative determinant.
\begin{align}
\textrm{det}\left(\textbf{V}-\frac{\mathbb{I}}{2}\right ) = \begin{vmatrix}
\bar{n} & m\\ 
m^* & \bar{n}
\end{vmatrix} = \bar{n}^2-|m|^2 \geq 0\,.
\end{align}

It is important to emphasise that $\bar{n}$ is not necessarily the mean number of thermal photons $\bar{n}_{\textrm{th}}$, but the mean number of photons corresponds to the distribution in which the system is described. 
\end{document}